\documentclass[twocolumn]{aastex7}

\usepackage{graphicx}
\usepackage{txfonts}
\usepackage[]{hyperref}

\hypersetup{
    colorlinks = true,
    linkcolor = {blue}, 
    citecolor = {blue},
    urlcolor = {blue}
}

\usepackage{natbib}
\usepackage{color}
\usepackage{float}
\usepackage{xspace}

\newcommand{\teff}{\ensuremath{T_{\mbox{\scriptsize eff}}}\xspace}

\newcommand{\rprime}{\ensuremath{\log R^\prime_\mathrm{HK}}\xspace}
\newcommand{\smwo}{\ensuremath{S_{\mbox{\scriptsize MWO}}}\xspace}
\newcommand{\sapo}{\ensuremath{S_{\mbox{\scriptsize APO}}}\xspace}
\newcommand{\BpRp}{\ensuremath{G_{\mbox{\scriptsize BP}}- G_{\mbox{\scriptsize RP}}}\xspace}
\newcommand{\hk}{\ion{Ca}{2} \text{H \& K}\xspace}

\begin{document}

\title{OWLS I: The Olin Wilson Legacy Survey}

\author[0000-0003-2528-3409]{Brett M.~Morris}
\affiliation{Space Telescope Science Institute, 3700 San Martin Dr, Baltimore, MD 21218, USA}
\email{bmmorris@stsci.edu}

\author[0000-0003-1263-8637]{Leslie Hebb}
\affiliation{Physics Department, Hobart and William Smith Colleges, 
300 Pulteney St., 
Geneva, NY 14456, USA}
\affiliation{Department of Astronomy and Carl Sagan Institute, Cornell University, 122 Sciences Drive, Ithaca, NY 14853, USA}
\email{hebb@hws.edu}

\author[0000-0002-6629-4182]{Suzanne L. Hawley}
\affiliation{Department of Astronomy, Box 351580, University of Washington, 
Seattle, WA 98195, USA
}\email{slhawley@uw.edu}

\author[0000-0002-2316-6850]{Kathryn Jones}
\affiliation{Center for Space and Habitability, Gesellsschaftstrasse 6, 3012 Bern, Switzerland}
\email{kathryn.jones@unibe.ch}

\author{Jake Romney}
\affiliation{Physics Department, Hobart and William Smith Colleges, 
300 Pulteney St., 
Geneva, NY 14456, USA}
\email{jake.romney@hws.edu}



 
\begin{abstract}

We present initial results from a planned 10~year survey of Ca II H \& K
emission, using observations made with the ARC 3.5m Telescope at Apache
Point Observatory.  The primary goal of the survey is to investigate
activity cycles in low mass stars.  The sample includes stars chosen from
the legacy Mount Wilson survey carried out by Olin Wilson more than 50 years ago,
together with newly identified planet-host stars and a select sample of early-mid M~dwarfs.
This paper presents the first four years of data, comprising 1040 observations
of 271 stars, with a specific focus on K and M stars.  We identify a
subsample of 153 stars for continuing observations over the full
10~year survey. Early results indicate that our data are
consistent with the MWO cycle periods over a time span of more than 50
years; that there is a bifurcation in activity in the late K range with
separate populations of low and high activity stars at lower masses; and
that M dwarf planet hosts tend to be mainly found in the population of low
activity stars, even in the unbiased (by activity) TESS sample,
potentially indicating a link between activity and planet formation.
We have also found indications of possible cyclic variability in some
of the lower mass stars in the sample.  Our ultimate goal is
to link the activity cycle and rotation periods in a robust sample of
stars spanning FGKM spectral types and to investigate the implications
for the underlying magnetic dynamo.
\end{abstract}


\section{Introduction}\label{sec:intro}

Magnetic activity in the photospheres of stars like the Sun manifests as dark spots in the optical, where strong magnetic fields locally inhibit convection, cooling roughly planet-sized regions in the photosphere \citep{Solanki2003}.  Coverage by active regions varies as spots emerge and decay (over timescales of months), as spots come in and out of view as the star rotates (from days to months), and with the phase of the stellar activity cycle (typically years).  While these active regions may present as dark spots in broad optical bandpasses, they emit excess flux compared to the mean solar disk in emission lines such as the resonant optical transitions of the first ionized state of Calcium, called H \& K by \citet{Fraunhofer1817}.  These are the strongest emission lines emitted by the solar chromosphere that are visible with ground-based telescopes. The discovery of a direct correlation between \ion{Ca}{2} K line emission and magnetic field strength in solar plage \citep{Leighton1959} provided early empirical evidence for the use of \hk emission as a proxy for stellar magnetic field activity. 

The emission in the \hk lines probes the upper photosphere and lower chromosphere of the star. The shape of the features reflects the temperature and density at increasing heights of the stellar atmosphere with distance from the core of the lines. In particular, the decrease in temperature with height through the outer layers of the photosphere produces a deep absorption line in the wings until the temperature reversal at the base of the chromosphere. The increase in temperature with height through the lower chromosphere produces an emission feature in the central 1~\AA\ of the line \citep[][and references therein]{Cretignier2024}.  The hotter the chromosphere relative to the photosphere, the larger the emission feature in the center of the \hk lines.  Magnetically active regions on the star which heat the chromosphere produce the excess emission that is observed in disk-integrated spectra of active stars.  Since the 1950s, many scientific programs have used observations of \hk emission as a proxy for magnetic activity to study the strength and evolution of stellar magnetic fields and how magnetic activity relates to fundamental stellar properties, such as mass, metallicity, spectral type, rotation period, and age \citep[e.g.][]{Vaughan1978,Linsky1979,Noyes1984,Schrijver1987}.  In addition, much work has been done to calibrate equivalent \hk measurements of the Sun in order to place the Sun in the stellar context \citep[e.g.][]{Hall2007,Egeland2017,Lorenzo-Oliveira2018}.

Olin C.~Wilson was the first to measure long-term, periodic activity cycles similar to the Sun's 11-year cycle in solar analogue stars \citep{Wilson1957,Wilson1963,Wilson1964,Wilson1968,Wilson1970,Wilson1976,Wilson1978}.  Over the span of three decades, he and collaborators (Art Vaughan, George Preston, Douglas Duncan, Sallie Baliunas, and many others) collected a treasure trove of spectroscopic observations of the \hk fluxes of bright solar-type stars to explore numerous aspects of stellar magnetic activity.  This seminal observing program began in 1966 on the 60~inch telescope at Mount Wilson Observatory (MWO)\footnote{Mount Wilson Observatory is not named for O.C.~Wilson.}, and has become known as the Mount Wilson Observatory HK~project.  By 1980, the 60~inch was completely dedicated to this project and continued to collect \hk fluxes on bright nearby stars until 2002 \citep{HK_project}.  The nature of the data collected by the photoelectric spectrometer, which uses a single photomultiplier to detect four channels of spectral data centered on the H and K line cores as well as two bracketing continuum regions ($R$ and $V$), necessitated the conversion of a star's \hk flux into a single spectroscopic index, called the S-index \citep{Duncan1991}.  The last complete and calibrated data release of MWO S-index values occurred in 1995 and includes over 300,000 observations of 2288 stars \citep[][]{Baliunas1995, HK_project}\footnote{The MWO archival observations are publicly available at: \url{https://dataverse.harvard.edu/dataverse/mwo_hk_project}\label{dataverseMWO}}.  While the majority of the stars in the sample have less than 10 observations, there are 91 stars that have more than 1000 independent activity measures spanning 30 years between 1966-1995.  Several foundational studies have been based on these data, including: the relationship between spectroscopic activity indicators and stellar rotation \citep{Skumanich1972,Barnes2007,Mamajek2008}; the effects of magnetic activity on radial velocity time series \citep{Saar1997,Lovis2011}; and the relationship between magnetic activity and stellar dynamos \citep{Radick1998,Bohm-Vitense2007,BoroSaikia2018}.

After the discovery of long-term activity cycles in solar analogues by \citet{Wilson1978},  \citet{Baliunas1995} and \citet{Radick1998} used the expanded dataset to reveal the diversity of stellar activity cycle morphologies which vary as a function of stellar age and mass and thereby defined four different activity types: variable, cyclic, trend, and flat activity. Stars like the Sun (G, early K) may have regular activity cycles with periods of around a decade, whereas young solar analogs often have chaotic, non-periodic ``cycles'', and higher mass (early G, F) stars may show no detectable long-term variations at all.  There were not enough late K and M stars in the sample to draw signficant conclusions about their cyclic activity. Since the late 1990s, time series spectroscopic data have often been acquired for the purposes of detecting extra-solar planets, these data have been used to further explore long-term activity cycles in some of the MWO stars and in new samples of lower mass and/or planet hosting stars using the \hk feature \citep[e.g.][]{Wright2004,Hall2007, Isaacson2010, Arriagada2011,Lovis2011,Luhn2020,GomesdaSilva2021,Fuhrmeister2023, Mignon2023,Isaacson2024}.

Here we describe the Olin Wilson Legacy Survey (OWLS), a new long-term \hk activity survey using the ARC 3.5m Telescope at Apache Point Observatory and the ARCES echelle spectrograph. We have several goals for our survey.  First, we will measure optical spectra including \hk and identify any long-term activity cycles that are present in a large number of recently discovered, planet-host stars.  These observations will be used to characterize the magnetic activity on planet-host stars and its variation in time, which will be useful for investigating the effects of activity on planetary atmospheres. 

Early planet-host transit surveys concentrated on finding planets around sun-like G stars (e.g. Kepler) which turned out to be difficult because the habitable zones are relatively far away, making transits infrequent. There has been significant recent interest in searching for earth-like planets around late M dwarfs where the habitable zones are very close (e.g. Trappist 1).  However, the high X-ray luminosities and frequent flaring exhibited by most of these stars may strip the atmospheres of potentially habitable planets, making them less suitable for life than earlier-type host stars \citep{Shkolnik2014,Owen2016,Airapetian2017,GarciaSage2017}. Therefore, the OWLS sample focuses on K dwarf planet hosts, where earth-like planets are still detectable in photometric surveys from space, but the habitable zones lie farther from their host stars than in late M dwarfs, and the typical activity levels are lower.  In essence, this is an activity level framing of the ``K dwarf opportunity" that is identified in recent papers outlining their potential advantage over both G-dwarfs and M-dwarfs in the search for habitable exoplanets and life on other worlds \citep[e.g.][]{Cuntz2016,Arney2019,Richey-Yowell2019,LilloBox2022,Hatalova2023,Vilovic2024,Richey-Yowell2023}.  

Second, we will observe a legacy sample of FGK stars that were observed as part of the original MWO project to determine if periodic activity cycles observed 30-60 years ago are still continuing.  We will further constrain the observed periodic variability and explore whether the nature of the activity cycles has changed over time.  For example, our data will be used to identify whether any chaotically variable stars have become periodic or whether previously variable objects have transitioned into an inactive state equivalent to a ``Maunder minimum".

Finally, we will investigate possible activity cycles in a sample of active early to mid~M~dwarf stars that were previously (with smaller telescopes) too faint to observe at \hk and are not usually included in planet searches due to their known activity.  Our observations will connect \hk activity to previous and ongoing H$\alpha$, UV and X-ray surveys in active M stars, and will extend our knowledge of activity cycles into the M dwarf regime.

Ultimately, we hope to obtain cycle periods for a large number of stars that span the FGKM spectral types, and to compare these with the stellar rotation periods to gain insight into the operation of the underlying magnetic dynamo \citep{Bohm-Vitense2007}.

We begin with a description of the OWLS sample selection and statistics \S\ref{sec:sample}.  In \S\ref{sec:methods}, we describe the data collection and the S-index derivation.  Finally, in \S\ref{sec:results} and \S\ref{sec:future}, we present preliminary results and future work, respectively. 

\section{Sample}
\label{sec:sample}

Our investigation of stellar activity cycles has two phases.  In this paper (phase 1), we report on the entire sample of 271 stars (see Table~\ref{tab:maintable}), summarizing the data from the first four years of observing.  We also describe the continuing sample of 153 stars that will be monitored for the full 10 years of the planned survey (phase 2).

The full OWLS sample reported here comprises 271 stars selected from several sources. We first queried the NASA Exoplanet archive in spring 2020 for all stars with at least one planet found by transiting methods that are above a declination of $-10^\circ$ and within the magnitude limits of $3 < G < 12$ to match the capabiilities of our observing system (ARC 3.5m Telescope and ARCES echelle spectrograph).  This initial planet-host sample of more than 500 stars includes objects from the Kepler \citep{Kepler2010,KOIs2018} and K2 \citep{Kruse2019} surveys, as well as stars from WASP \citep{Pollacco2006}, HAT \citep{bakos2004,bakos2010}, and other ground-based surveys.  It also includes all of the 51 TESS stars \citep{TESS2015} identified as TOIs \citep[TESS Objects of Interest]{TOIs2021} at the time of survey start (2020) and accessible from APO.  From this initial sample, we observed 163 objects, concentrating primarily on brighter ($G<11.5$) and later-type (K-M) stars. All of these stars are denoted with ``1'' in the ``Planet'' column of Table~\ref{tab:maintable}. Note that all of the 51 TOIs in the initial sample were observed, and that some of the TOIs are still planet-host candidates, rather than confirmed planet-host stars.

As a legacy survey, we included dozens of stars from the original MWO sample \citep{Baliunas1995,Duncan1991}, with varying activity level, and with evidence for either cyclic (periodic) or chaotic activity. We concentrated mostly on the K dwarfs in the MWO sample, but did also include some F and G stars and the handful of M dwarfs that they observed.  If an MWO star was later found to have a planet, this is noted in the Planet column of Table~\ref{tab:maintable}. There are 13 such MWO stars that are now known to have planets. The OWLS sample thus includes a total of 176 planet-host stars, denoted with ``1'' in the ``Planet'' column of Table~\ref{tab:maintable}.

Finally, to augment the number of stars of later spectral types, we chose a few dozen late-K and early-M stars from The Palomar/MSU (PMSU) Nearby Star Spectroscopic Survey of nearby low-mass stars \citep{PMSU1,PMSU2}, as well as some well-known active mid-M dwarfs such as AD Leo, EV Lac and YZ CMi.

In the final 271 star sample that we report here, there are 79 MWO stars (mostly K) of which 13 have planets; 163 stars with planets found by transit methods (mostly K); and 29 stars from the PMSU survey (late K and M type) that do not have known planets.  There are 176 total stars with planets in the full sample (Table~\ref{tab:maintable}) and 95 stars without known planets (see column 3).

\begin{deluxetable*}{lccccccccccccc}[h]
\tablenum{1}
\tablecaption{Full sample}
\label{tab:maintable}
\centering
\tablehead{\colhead{Name} & \colhead{Sample} & \colhead{Planet} & \colhead{$G$} & \colhead{\BpRp} & \colhead{Parallax} & \colhead{M$_{G}$} & \colhead{Spectral} & 
\colhead{Monitoring} & \colhead{Mean} & \colhead{Num} & \colhead{$B-V$} & \colhead{\teff} & \colhead{\rprime}  \\[-0.2cm]
\colhead{} & \colhead{} & \colhead{} & \colhead{(mag)} & \colhead{(mag)} & \colhead{(mas)} & \colhead{} & \colhead{Type} & \colhead{Sample} & \colhead{\smwo} & \colhead{Obs} &
\colhead{(mag)} & \colhead{ (K)} & \colhead{}
}
\startdata
\hline\hline
HD 224930 &  MWO & 0 & 5.50	& 0.92 & 81.2	& 5.05 & G & 0 & 0.178 & 1 & 0.748 &5440 &-4.747\\
HD 3167 &  planet & 1 & 8.76 & 0.99 & 21.1 & 5.39 & K & 1 & 0.175 &5 &0.827 &5244 &-4.799 \\
HD 3443 &  MWO & 0 & 5.95 & 0.92 & 64.9 & 5.02	& G & 0 & 0.170 & 3 &0.748 &5440 & -4.783 \\
WASP-93 &  planet & 1 & 10.98 & 0.63 & 2.7 & 3.11 & F & 0 & 0.195 & 2 & 0.477 & 6387 & -4.608\\
HAT-P-16 &  planet & 1 & 10.76 & 0.73 & 4.4 & 3.97 & F & 0 & 0.160 &1 &0.565 &6036 & -4.828 \\
HD 3651 &  MWO & 1 & 5.63 & 1.02 & 89.8 & 5.40 & K & 1 & 0.164 & 5 & 0.860 & 5162 & -4.870 \\
HD 3795 & MWO & 0 & 5.93 &0.93 & 35.1 & 3.66 & G & 0 & 0.156 & 3 & 0.757 & 5420 & -4.870 \\
GJ 29.1 & PMSU & 0 & 9.55 & 2.12 & 46.0 & 7.87 & M & 0 & 5.422 & 1 & 1.493 & 3630 & -3.502 \\
... & & & & & & & & & & & & & \\
 HD 22049 &  MWO & 1 & 3.37 & 1.12 & 310.6 & 5.83 & K & 1 & 0.456 & 9 & 0.903 & 5050 & -4.247 \\
... & & & & & & & &  & & & & & \\
HAT-P-6 & planet & 1 & 10.31 & 0.59 & 3.6 & 3.09 & F & 0 & 0.177 & 2 & 0.443 & 6538 & -4.722
\enddata
\end{deluxetable*}

Table~\ref{tab:maintable} lists data for the full sample, with a ``1" in column 8 denoting the objects in the phase~2 ``monitoring sample'', that will continue to be observed for the full 10 years of the survey.  The stars are listed in order by Right Ascension, although the coordinates are not shown.  The Gaia magnitude and color, G and (\BpRp), and the Gaia parallax and absolute magnitudes were obtained or calculated from Gaia DR3 \citep{gaiaDR3}.  The spectral types and effective temperatures were obtained from the \BpRp color using the relationships found in an updated version of Table~5 from \citet{Pecaut2013}\footnote{The most recent version is from 2024 and publicly available at: \url{https://github.com/emamajek/SpectralType/}\label{mamajektable}}.  As described further in Section~\ref{sec:methods} below, the S-index on the MWO scale, \smwo , is calculated from standard stars obtained using the APO observing system, and the \rprime \citep{Noyes1984,Vaughan1978} is calculated using the B-V and effective temperatures listed here, which are derived from the \BpRp color.  Table~\ref{tab:maintable} thus provides a fully consistent set of data, including measured \hk indices for each object in the sample, with derived quantities based only on the APO observations and the measured Gaia \BpRp color.

Table~\ref{tab:spectype} provides some statistics on the breakdown of the sample by color and spectral type.  As described above, we have concentrated on K stars, and the 10 year, phase 2 monitoring sample clearly reflects this goal, with K stars comprising nearly 2/3 of the sample.  The size of the monitoring sample was determined by the availability of telescope time, and the desire to observe each star at least once, and preferably a few times, per year for the 10 year duration of the survey.  Particularly for the redder stars, the monitoring sample includes mostly the brighter objects which provide better S/N in our 30 minute exposure times.
\begin{figure}[h]
    \includegraphics[width=\columnwidth]{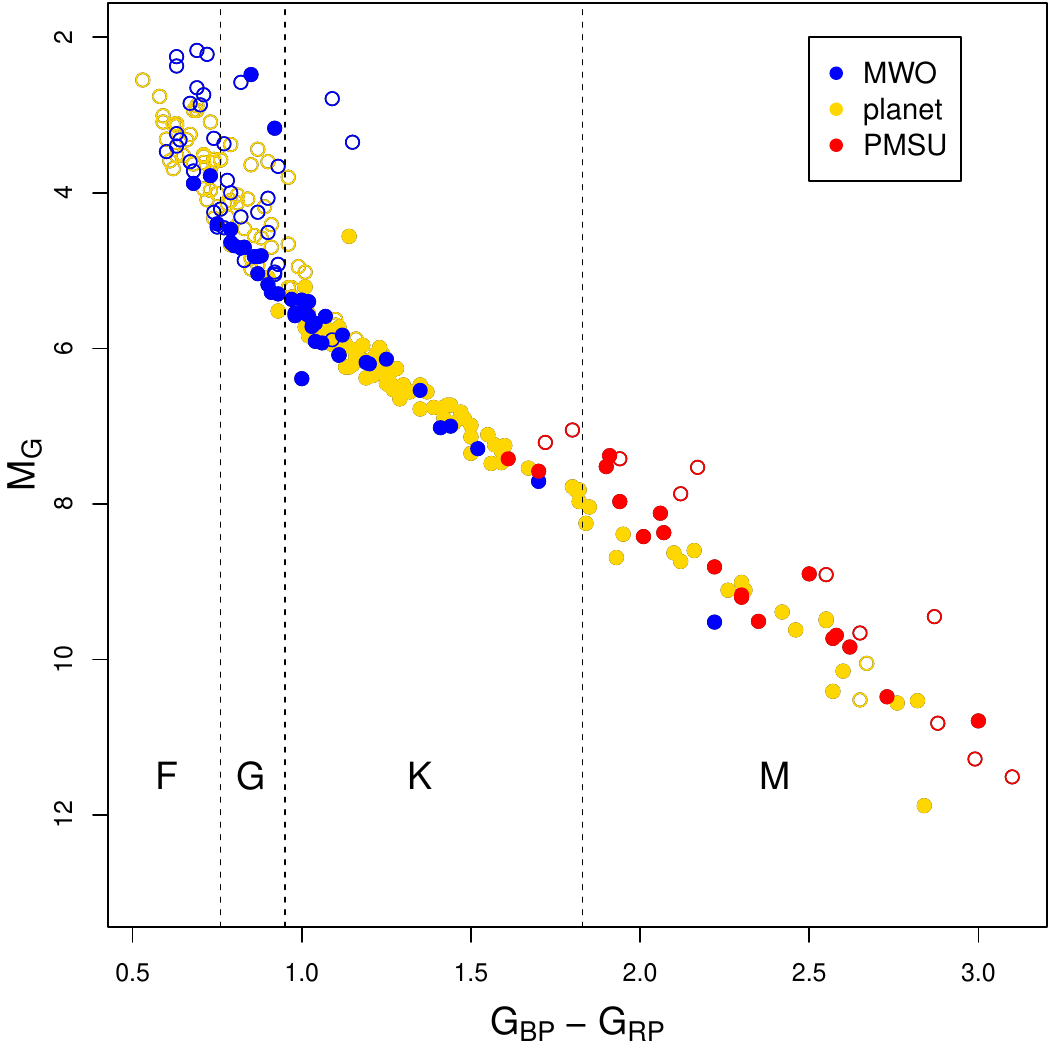}
    \caption{Color-magnitude diagram of the full OWLS sample using Gaia photometry.
    Blue points are stars observed in the original Mount Wilson survey (MWO), gold points are known to be planet-hosting stars, and red points are late type stars without known planets from the PMSU survey.  The closed symbols represent stars that we retain in the ongoing 10-year monitoring sample.}
    \label{fig:cmd}
\end{figure}

Figure~\ref{fig:cmd} shows a Gaia color-magnitude diagram of OWLS targets. The three samples (MWO, planet, PMSU) are indicated by the blue, gold and red points respectively, with the solid symbols showing the 153 stars that will continue to be monitored for the 10 year survey (see also column 8 in Table~\ref{tab:maintable}).  Many of the stars that lie above the main sequence are subgiants (F and G stars) and/or binaries (K and M stars), and are generally excluded from the monitoring sample.

\begin{deluxetable}{cccc}
\tablenum{2}
\tablecaption{Sample size by spectral type}
\label{tab:spectype}
\centering
\tablehead{\colhead{\BpRp} & \colhead{Spectral} & \colhead{Full} & \colhead{Monitoring}  \\[-0.2cm]
\colhead{} & \colhead{Type} & \colhead{Sample} & \colhead{Sample} 
}
\startdata
\hline\hline
$\le 0.76$ & F	& 53 & 3  \\
$0.77-0.95$ & G	& 50 & 15  \\
$0.96-1.83$ & K	& 121 & 99  \\
$\ge 1.84$ & M	& 47 & 36  \\
\hline
 & & 271 & 153
\enddata
    
\end{deluxetable}

\section{Methods} \label{sec:methods}

The S-index is defined for a given observatory/instrument configuration, and then a simple linear transformation ``calibrates'' the S-index from a given observatory to a common scale. The large existing public database of the S-index for bright northern stars from Mount Wilson Observatory
\citep{Baliunas1995,Duncan1991} serve as the benchmark observations, \smwo,  against which new observations at other observatories are calibrated. 
In \citet{Morris2017b, Morris2019b}, we observed numerous MWO stars with the ARCES echelle spectrograph to calibrate the APO/ARCES observing system, enabling us to measure \smwo of any star observable with APO/ARCES.

\subsection{APO/ARCES Spectra}

OWLS survey observations are being obtained with the ARCES echelle spectrograph on the Astrophysical Research Consortium (ARC) 3.5 m Telescope at Apache Point Observatory. ARCES is a $R\sim31{,}500$ spectrograph that has been permanently mounted on the $f/10$ Nasmyth 1 port of the ARC 3.5 m Telescope since 1999.  Observations span optical wavelengths from $0.32-1~ \mu$m.   During each night of observations, we collect bias frames, ``red'' and ``blue'' flats with the internal calibration quartz lamp without and with a blue filter in place, respectively, and Thorium-Argon lamps for wavelength calibration.

We process the raw ARCES spectra with \texttt{IRAF} methods to subtract the bias, remove cosmic rays, normalize by the flat field, trace and extract all echelle orders, and then perform the wavelength calibration based on the standard ARCES data reduction manual\footnote{An ARCES data reduction manual by J.~Thorburn is available at \url{https://www.apo.nmsu.edu/arc35m/Instruments/ARCES/images/echelle_data_reduction_guide.pdf}}. The reduced spectra will be publicly available\footnote{\url{https://archive.stsci.edu/hlsp/owls/}} on the Mikulski Archive for Space Telescopes (MAST) as a High Level Science Product (\dataset[10.17909/vet0-jg24]{\doi{10.17909/vet0-jg24}}), and we provide an interactive visualization dashboard for OWLS observations in an open source Python package \footnote{\url{https://github.com/bmorris3/owls-app}}.

After the basic data reduction, trace and order extraction, and wavelength calibration is complete, we remove the blaze function of all orders of each science spectrum by dividing by a high-order polynomial fit to the corresponding orders of the spectrophotometric standard star, HZ~44 \citep{Massey1988, Oke1990, Landolt2007} obtained on 2024-June-19. 

The normalized spectra are then shifted in wavelength into the rest-frame by correcting for the barycentric motion of the Earth at the time of the observation and the stars' radial velocities.  We remove the wavelength shift by maximizing the cross-correlation of the ARCES spectra compared with a PHOENIX model atmosphere spectra \citep{Husser2013} matched to the temperature of the science target.  We use 53 echelle orders with spectral features from 4000-6500\AA\ in the cross-correlation to find a single velocity value for each spectrum.  We then apply the wavelength shift to each spectral order based on the single velocity value and the median wavelength of that order.  

\begin{figure}
    \centering
     \includegraphics[width=\columnwidth]{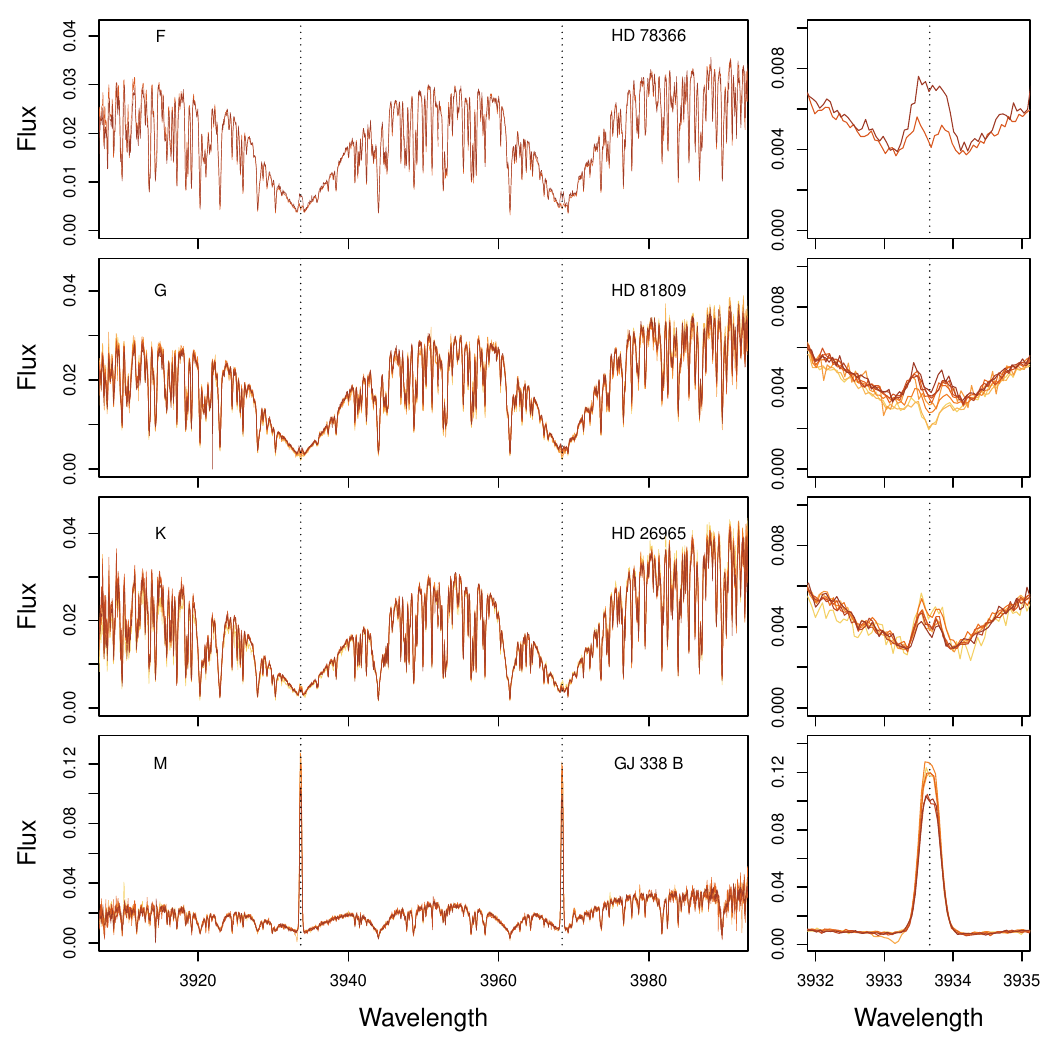}
     \caption{{\bf Left:} All available spectra obtained from the APO ARCES instrument between June 2020-July 2024 for four representative targets in the wavelength region from 3910-3990~\AA.  Two spectra of the F~dwarf HD~78388, eight spectra of the G~dwarf HD~81809, seven spectra of the K~dwarf HD~26965, and six spectra of the M~dwarf GJ 338~B are shown from top to bottom of the figure.  Each spectrum is scaled so the continuum and absorption features align over the entire wavelength region.  {\bf Right:} Variations in the emission line core are visible in the right-hand plots that highlight the \ion{Ca}{2}~\text{K} feature.}
    \label{fig:spectra}
\end{figure}

Figure ~\ref{fig:spectra} presents all available APO spectra for four representative targets in the wavelength region from 3910-3990~\AA.  The complete \hk region shown is concatenated from two consecutive orders at 3952~\AA\ (trimming the low-signal-to-noise edges of each order) and scaled to a common value for all spectra by dividing by the integrated flux values from the nearby 20\AA-wide pseudocontinuum regions $R$ and $V$ centered on 3900 and 4000~\AA, respectively.  The continuum and the absorption features match for all observations of the same target due to this scaling, but there are visible differences in the line cores due to the changing magnetic activity that are reflected in the measured S-index values.  

\subsection{Measuring the S-index}

Following \citet{Baliunas1995}, we measure the S-index for each star on the APO scale by taking the sum of the flux in the cores of the $H$ and $K$ features at 3968.47 \AA\ and 3933.66 \AA, weighted by a triangular weighting function with FWHM = $1.09$ \AA. The weighted emission is normalized by the flux in the pseudocontinuum regions $R$ and $V$, which are 20 \AA-wide bands centered on 3901 and 4001 \AA, respectively. The summed flux values are combined in the following way to provide the S-index on the APO scale
\begin{eqnarray}
S_\mathrm{APO} &=& \frac{a~H + b~K}{c~R + d~V}, 
\end{eqnarray}
where $a,~b,~c,~d$ are parameters that can be tuned so that $S$ has roughly equal flux contributions from the $H$ and $K$ emission lines, and roughly equal flux from the $R$ and $V$ psuedocontinuum regions in the APO spectra \citep{Duncan1991,Isaacson2010}. 
 Here, we set $a = c = 1$, and we choose $b=0.84$ and $d=1$, so that $H \sim b~K$ and $K \sim d~V$. 

We then place the APO S-index, \sapo, on the MWO-calibrated scale according to the following equation
\begin{eqnarray}
\smwo &=& C_1 S_\mathrm{APO} + C_2, \label{eqn:s_ind}
\end{eqnarray}
where $C_1$ and $C_2$ are deterimined as follows, to make the $S$-indices derived from APO spectra match the scale of \smwo \citep{Duncan1991}.

\begin{figure}
    \centering
    \includegraphics[width=\columnwidth]{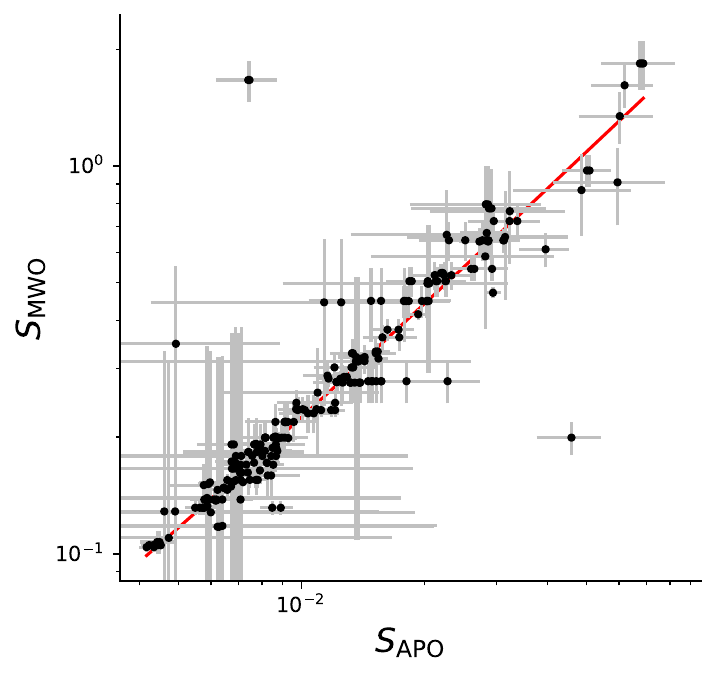}
    \caption{APO S-index values compared to MWO S-index values for the same 84 calibration stars.  The red line is a linear fit to the measurements with $C_1=20.9_{-0.5}^{+0.6}$ and $C_2 = - 0.012 \pm 0.003$.  This is a recalibration of the MWO S-index using the ARCES spectra that was presented in \citet{Morris2017b} with only 30 calibration stars.}
    \label{fig:s-recalibration}
\end{figure}

We calibrate the ARCES spectra following the procedure developed in \citet{Isaacson2010} for Keck/HIRES and applied to APO/ARCES data in \citet{Morris2017b} with 30 calibration stars.  Here, we measure the \sapo from APO/ARCES spectra of 84 known MWO targets with a range of activity levels.  These values are compared to \smwo values from \citet{Baliunas1995} and \citet{Duncan1991} obtained via the Harvard Dataverse\textsuperscript{\ref{dataverseMWO}}.   Since $S$ varies over time for each star in the sample, the linear correlation between the $S_\mathrm{APO}$ and \smwo will have some intrinsic spread. To incorporate this into our model, we adopt the $\left< S \right>$ and the standard deviation of $S$ from \citet{Duncan1991} as the measurement and uncertainty of the MWO values. We solve for the updated constants $C_1=20.9_{-0.5}^{+0.6}$ and $C_2 = - 0.012 \pm 0.003$ with a Markov Chain Monte Carlo \citep{Goodman2010, Foreman-Mackey2013}.  The linear correlation we find between the measured $S_\mathrm{APO}$ values and the derived \smwo values is shown in Figure~\ref{fig:s-recalibration}.

\begin{figure}
    \centering
    \includegraphics[width=\columnwidth]{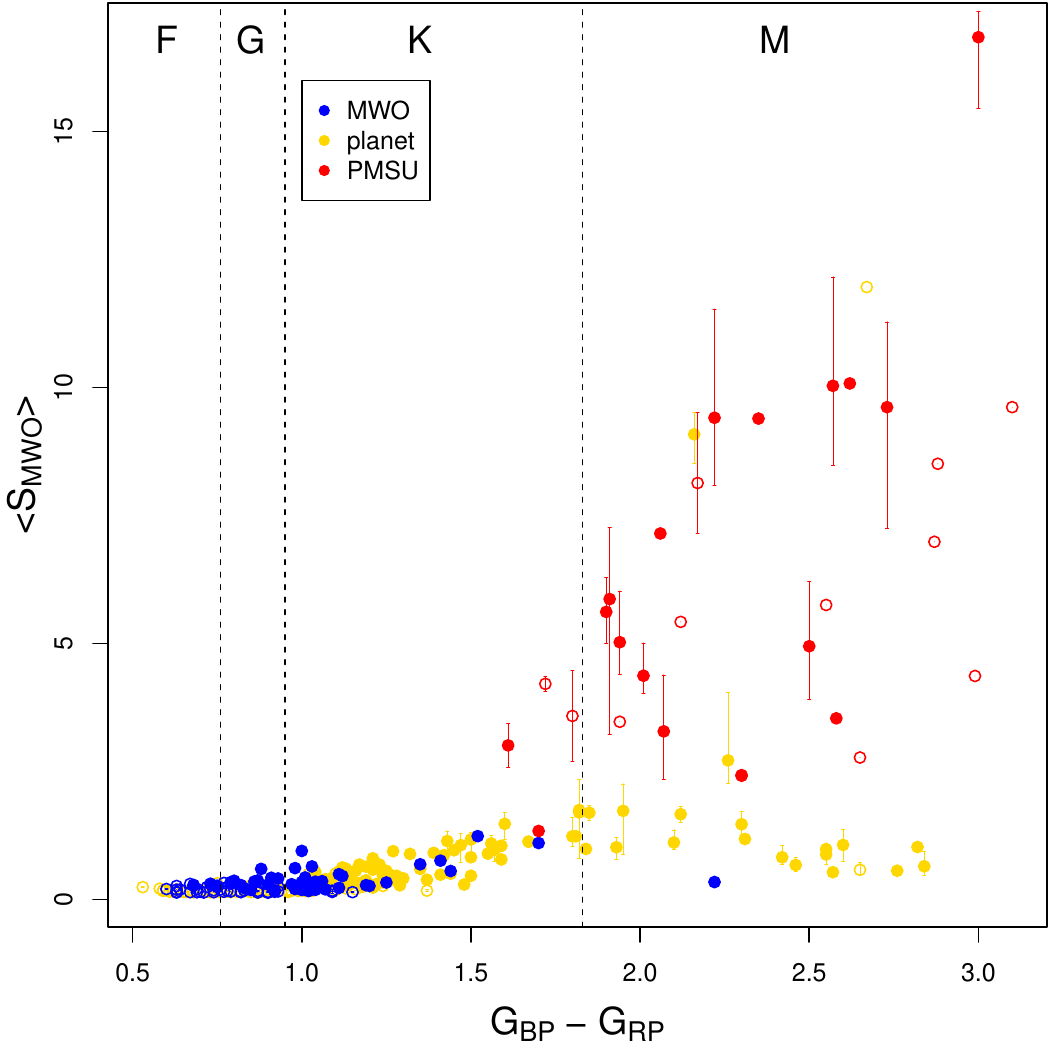}
    \caption{Mean \smwo values for the full sample are shown as a function of color (spectral type).  The point style is the same as in Figure~\ref{fig:cmd}: blue points are stars in the original Mount Wilson Survey, gold points are planet host stars, and red points are active M-dwarf stars from the PMSU survey.  The closed symbols represent stars that are included in the ongoing monitoring sample. The variation we measure in \smwo is indicated by the vertical bars for each star.}
    \label{fig:meanSmwo}
\end{figure}

All the \smwo values derived from the APO spectra of an individual object are averaged into a single mean \smwo value and presented in Column~9 of Table~\ref{tab:maintable} along with the number of APO spectra obtained on that object after four years of survey operations. The values are plotted in Figure~\ref{fig:meanSmwo} as a function of the Gaia DR3 \BpRp color.

\subsection{Calculating \rprime} \label{sec:rprime}

The method described above for calculating $S$-indices from observed spectra results in a value that includes contributions to the total 
\hk flux from both non-thermal chromospheric heating and thermal emission in the photosphere.   As a result, S-indices correlate with both magnetic activity and stellar temperature.  The \rprime statistic was developed by \citet{Linsky1979} to separate the true flux excess in the \hk line due to chromospheric emission from the photospheric contribution to the flux, allowing for a comparison between magnetic activity levels for stars of different spectral types. 

We follow the method of \citet{Mittag2013} to derive \rprime values from the mean \smwo values in our sample.  $ R^\prime_\mathrm{HK}$ is defined as the difference between the measured H \& K flux ($F_{HK}$) and the photospheric contribution to that flux ($F_{HK,\rm phot}$), normalized by the bolometric flux of the star ($\sigma T_{\rm eff}^4)$. The authors define empirical relationships between these quantities and a star's $B-V$ color, however, the $B-V$ measurements for the targets in our sample are often outdated, inconsistently derived, and less discriminating for the reddest objects.  To correct for these factors and still make use of the \citet{Mittag2013} empirical relations, we derive a proxy $B-V$ color and \teff value from Gaia DR3 \BpRp color for every star in our sample by linearly interpolating the most recent version of the empirical spectral type-color-temperature table from  
\citet{Pecaut2013}\textsuperscript{\ref{mamajektable}}.  
The interpolated $B-V$ colors and \teff values are provided in Table~\ref{tab:maintable}, columns 11-12.  While these proxy values are not measured quantities, they accurately reflect the underlying photometric and bolometric flux of each star in a uniform way, thus they serve the correct purpose in the \citet{Mittag2013} equations.  Therefore, we apply their Equations (2) and (4) to derive the $F_{HK}$ from the measured \smwo and the $B-V$ color; their Equations (20) and (21) give us the corresponding photospheric contribution to the integrated line flux ($F_{HK,\rm phot}$) from the $B-V$ color, and the normalizing bolometric flux is calculated directly from the interpolated \teff. These calculated quantities are combined in their Equation (24) into a uniformly derived \rprime for each measured \smwo and presented in the last column of Table~\ref{tab:maintable}.

\section{Results}
\label{sec:results}

\begin{figure}
    \includegraphics[width=\columnwidth]{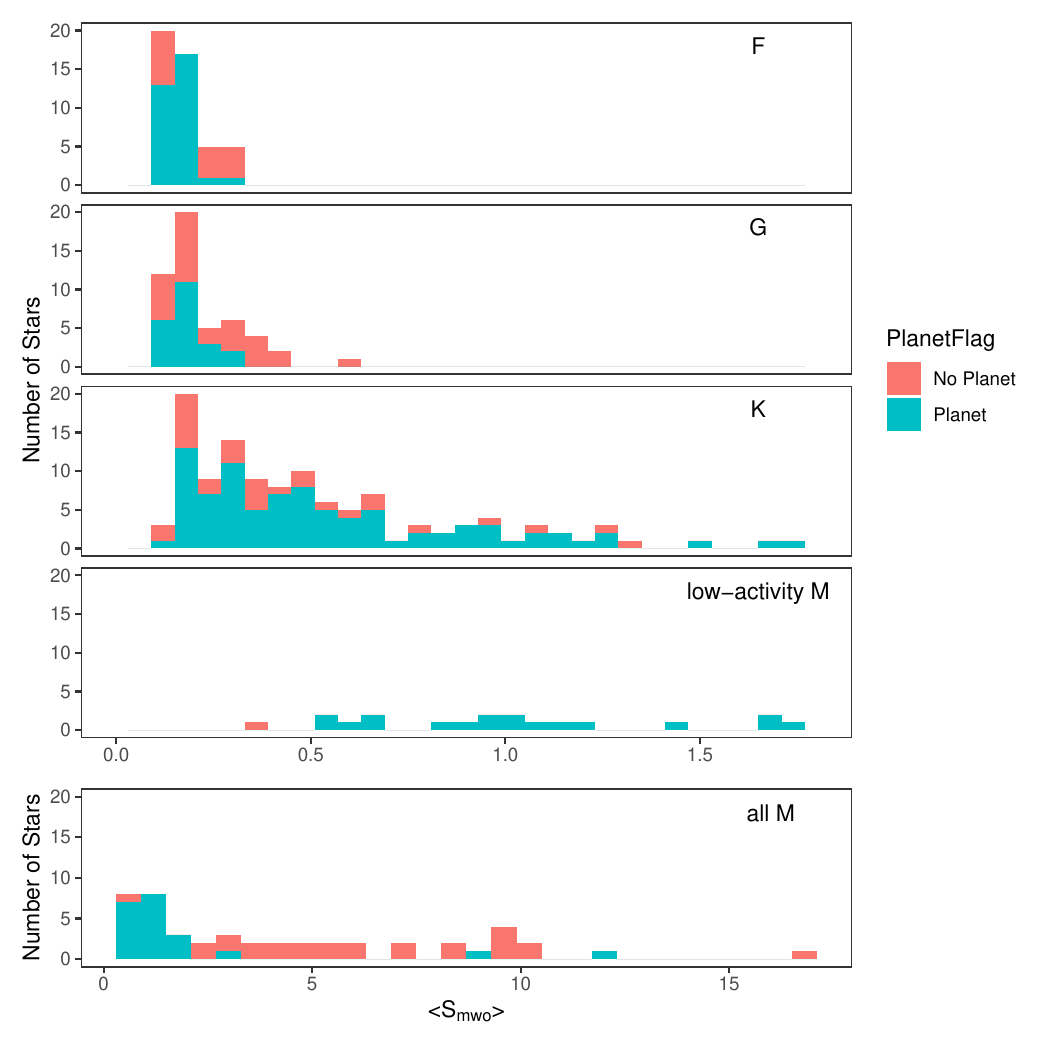}
    \caption{Histogram of \smwo for the full sample, divided into stars with planets (blue) and those without (orange). Increasing activity in the later type stars is evident. The second from bottom panel shows the low activity M dwarfs on the same scale as the higher mass stars, and the bottom panel is on a greatly expanded scale to include the high activity M dwarfs whose S-index is up to a factor of 10 higher. There is an intriguing indication that the highest activity M dwarfs preferentially do not have planets (bottom panel nearly all orange).}
    \label{fig:sindexhist}
\end{figure}

As described above, the \hk S-index and the corresponding \rprime  provide a way to compare the strength of the disk-integrated chromospheric emission in stars of different spectral types. Figures~\ref{fig:meanSmwo} and \ref{fig:sindexhist} show the distribution of mean \smwo values for the entire OWLS sample, as a function of \BpRp color and binned spectral type. In both figures, the earlier type F and G stars have relatively low values of \smwo of less than 0.5, with a modest spread.     For later type stars, Figure~\ref{fig:meanSmwo} shows an increase in average \smwo with increasing color throughout the K-dwarf range until around \BpRp = 1.5 (mid-late K stars) where the strength of the \hk emission begins to bifurcate, with some late-K and M-dwarf stars having low values, typical of early K stars, while others have much higher values. 
The histogram of activity values reflects this transition in the K-dwarf regime, showing both a significant fraction of low activity stars similar to F and G dwarfs and a tail of stars with larger \smwo values extending almost to 2.  The M star distribution is quite different with only a few stars below 0.5 (these are difficult to measure due to the reduced continuum and hence signal-to-noise in the spectra, but M stars with these low S-indices do exist), and many stars with very high S-index values up to nearly 20 as shown in the bottom panel with the expanded horizontal axis.  An example of such a high S-index star is given in Figure~\ref{fig:spectra} where the peak of the emission line is $\sim 20$~times higher for the M~dwarf, GJ~338~B, than the earlier F, G, and K-stars that are shown.  

Finally, according to Figure~\ref{fig:sindexhist} which compares the activity levels of planet to non-planet hosting stars, there does not appear to be an obvious difference in activity level between stars with planets (shown in blue) and those without (shown in orange) in the earlier spectral types (FGK).  However, the bottom panel shows that very few planets have been discovered orbiting the most active M stars. This may partly reflect a bias in radial velocity planet searches toward including stars with known low activity, but transit searches such as TESS should not be affected by such a bias.  It would be interesting to follow up on the effect of strong magnetic activity on planet formation around active M dwarfs. 

\begin{figure}
    \centering
    \includegraphics[width=\columnwidth]{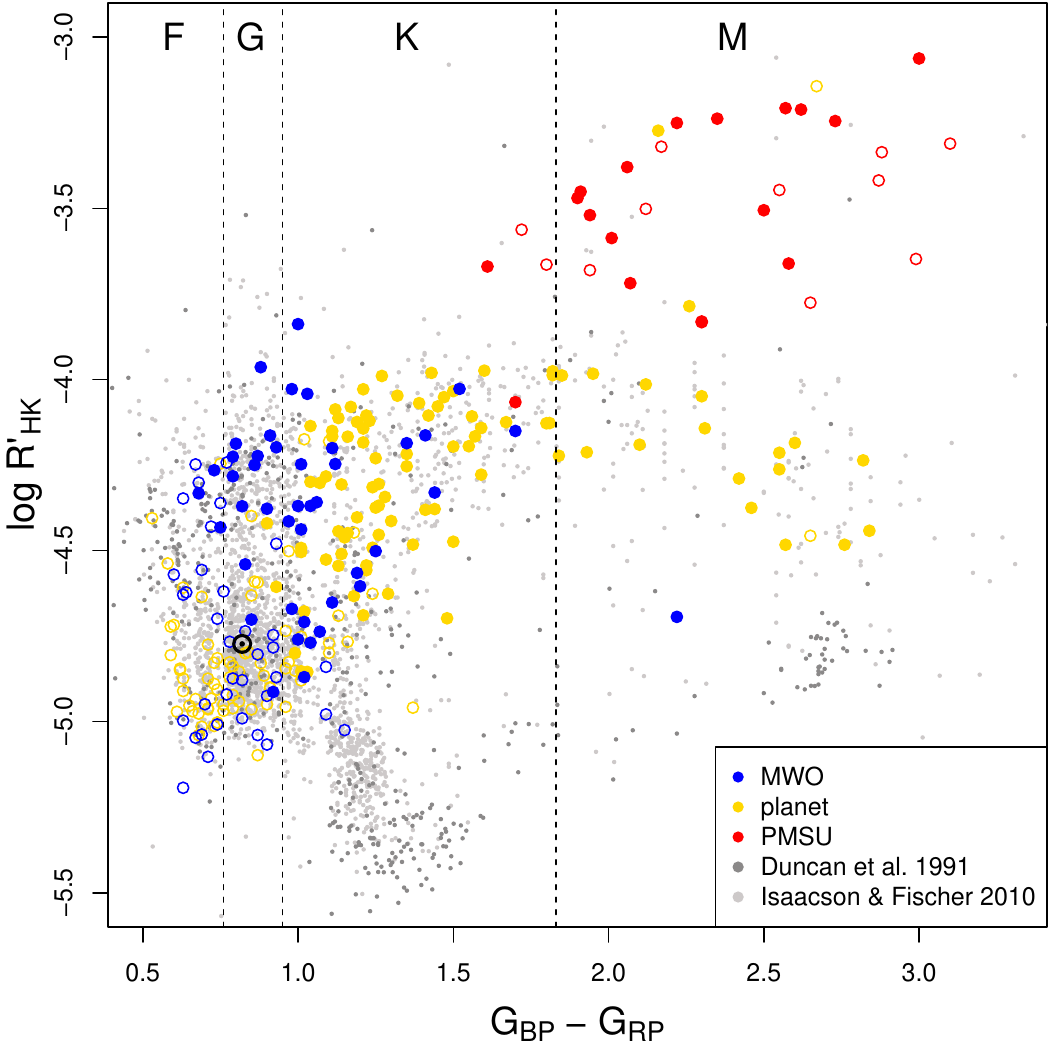}
    \caption{\rprime vs. \BpRp for OWLS targets (same color scheme as in previous figures) together with literature measurements (grey) from \citet{Duncan1991}, \citet{Isaacson2010}, and \citet{Wright2004}. The Sun is plotted in black with a \BpRp = 0.82 \citep{Casagrande2018} and \rprime = -4.77 derived in a consistent manner from the Sun's mean S-index value of 0.1694 \citep{Egeland2017}. }
    \label{fig:rhk}
\end{figure}

Figure~\ref{fig:rhk} shows \rprime versus \BpRp where the effects of the changing stellar continuum have been removed (see Section~\ref{sec:rprime}). The color scheme is consistent with earlier figures. The trends described in the \smwo data in Figure~\ref{fig:sindexhist} are even more evident in these data. There is a clear increase in \rprime values from F to G to K stars, and again the bifurcation into low and high activity stars is seen in the late K and especially in the M dwarfs.  In addition, the previously noted indication that high activity M dwarfs are much less likely to have planets is highlighted here by looking at the planet host stars (yellow symbols) all of which were found by transit surveys. Only a few of these stars have high activity, while the majority have low activity.  Interestingly, we observed all 51 TOIs that were identified at the time of survey start in 2020, without regard for, or any prior knowledge of, their activity status, and yet the vast majority have low activity.

To understand whether the bifurcation we see in the activity levels of the latest type stars in our sample is valid, we quantify the effect by fitting three different model distributions to a histogram of \rprime values for stars with \BpRp $> 1.5$: 
a Gaussian to describe a single population of stars with a preferred mean \rprime, a two Gaussian mixture model which represents two distinct populations of stars with different average \rprime values, and a flat distribution in which all \rprime values between -3.0 to -5.0 are equally likely.  Considering these three models, the double-Gaussian model has the maximum likelihood ($\mathcal{L}=-31$), and thus best describes the distribution of \rprime values for stars redder than \BpRp $>$ 1.5.  However, a more complex model with more free parameters will almost always have a better fit, so we also calculate the Bayesian Information Criterion (BIC) which accounts for the variation in complexity when comparing different models to the same data.  The double Gaussian also has the minimum BIC value of 83 even after accounting for its larger number of free parameters, confirming quantitatively that there are two separate activity distributions of stars in this color range.

We speculate on a few possible reasons for this bifurcation.  Despite choosing planet host stars only from transiting surveys, and without regard for their activity status, the planet-host sample may still be biased toward low activity stars if the transits are more easily identified in those stars.  The late type stars from the PMSU sample were chosen to be active, so they naturally sample the higher activity stars.  This may introduce an unintended effect when comparing the two samples.

If the bifurcation of activity is real, it may reflect similar effects seen in the rotation period distribution \citep{McQuillan2013} and in the period-age distribution \citep{Engle2023}.  In those cases, younger stars are rotating more rapidly, and this may be reflected in their magnetic activity.  Recently, \citet{Marvin2023} investigated \rprime in HARPS M~dwarfs and also found an apparent separation into higher and lower activity stars (see their Figure 8), with only the higher activity M dwarfs also showing H$\alpha$ emission.  A benefit of observing \hk is that the emission can be observed even in relatively inactive stars.  We intend to compare H$\alpha$ and \hk activity levels in our sample in future work.

The open and solid symbols depict the full and monitoring samples described in this paper.  We will continue to monitor both high and low activity late-K/M dwarfs to investigate whether cyclic activity is evident in either or both samples.  

The gray symbols show data from other large time series surveys of \hk \citep{Duncan1991,Isaacson2024}.  Of particular note is the population of K stars with very low activity levels (below \rprime = -5) in the other surveys.  These stars were chosen and monitored for planets because of their low activity. We do not include them in our sample because we chose stars with measurable activity in order to look for activity cycles.

\begin{figure*}
    \centering
    \includegraphics[width=0.9\textwidth]{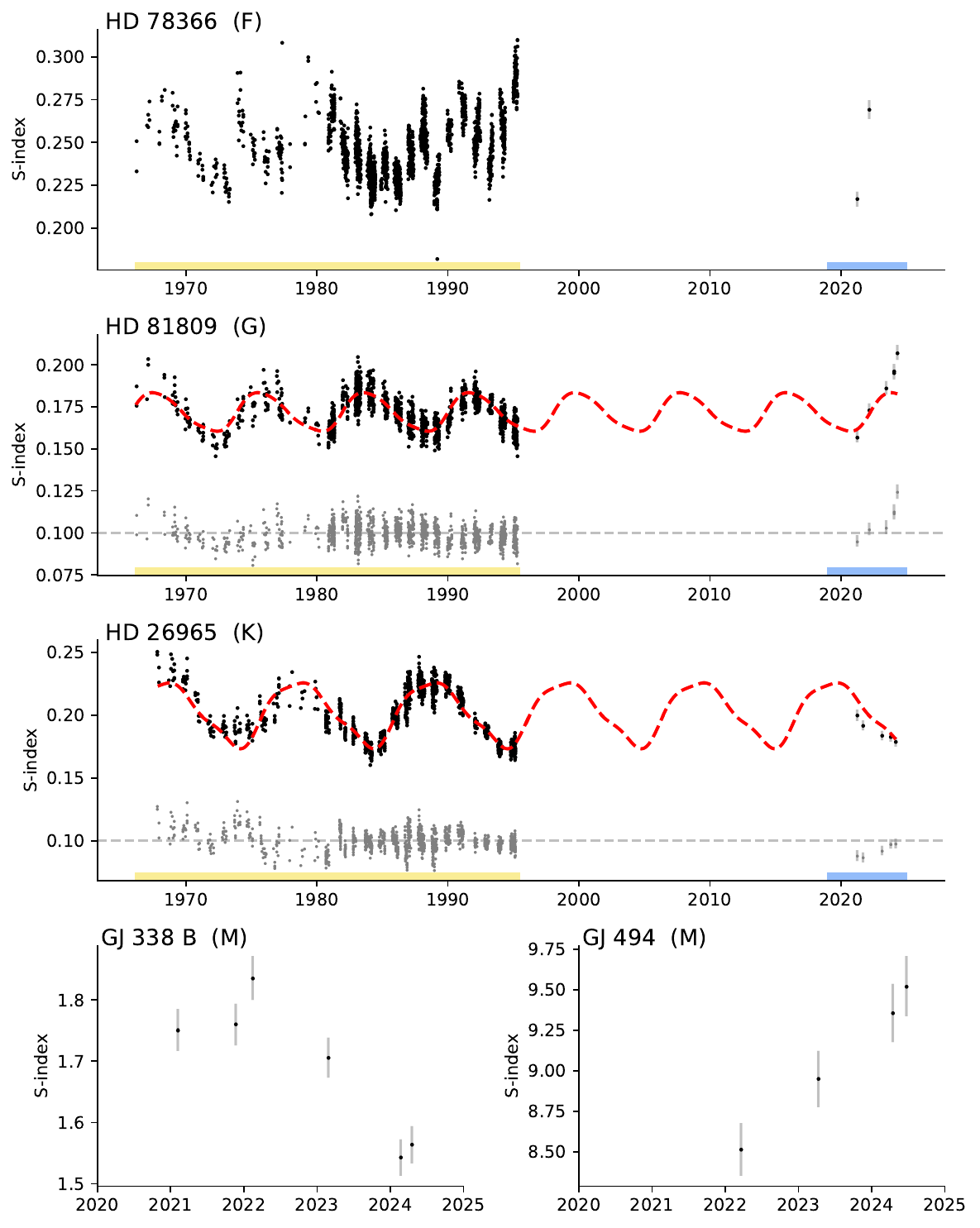}
    \caption{Time series of S-indices for five targets with a range of spectral types.  The top three panels show data from 1965 to 2024 for HD~78366 (F dwarf), HD~81809 (G dwarf), and the K-dwarf HD~26965, including historic MWO data (timespan in yellow) and newly obtained OWLS measurements (timespan in blue). We plot a curve on the G and K stars with three Lomb-Scargle (sinusoidal) terms, with periods ranging from 8 to 15 years. The OWLS S-indices from 2020-2024 for two M~dwarfs GJ~388~B and GJ~494 are shown in the bottom row.}
    \label{fig:lightcurves}
\end{figure*}

Figure~\ref{fig:lightcurves} shows example time series data (\smwo versus time) of the same stars for which we presented the individual spectra in Figure~\ref{fig:spectra}.  The F, G, and K-dwarf stars have more than 25 years of previous data from the MWO survey along with our newly derived S-index values from 2020-2024.  The data on the M-dwarfs are only from our survey over the past four years.  For the G and K stars, which appear periodic, we plot a curve to guide the eye.  The curve consists of three Lomb-Scargle (sinusoidal) terms, having periods ranging from 8 to 15 years, but our goal is to eventually carry out more sophisticated modeling of the periodic, cyclic behavior using the full 10-year survey data.  It is clear even with these simple fits to HD~81809 and HD~26965 that our data are consistent with the MWO observations indicating that the activity cycle periods have been stable for more than half a century.   HD~78366 is one of the three relatively active and variable F~stars that remain in our continuing sample.  It continues to show non-periodic variation with the same amplitude of variation as in the MWO data. The M stars both show suggestive variability over our four years of monitoring, but more data are needed to determine if they will exhibit periodic cycles. 

\section{Summary and Future Work}
\label{sec:future}

We have presented the first four years of data from a planned 10 year survey of \hk emission in FGKM stars that we have undertaken on the ARC 3.5m Telescope at Apache Point Observatory.  This paper provides activity measurements of 271 stars, of which 176 have planets.  For many of the planet-host stars, these are the first activity data that have been obtained.  They should thus be a valuable resource for characterizing the stellar environment for these newly discovered planets.  We also identified a monitoring sample of 153, primarily K and early M, stars that we will continue to observe for several more years.  

Preliminary results show that high activity M dwarfs may be less likely to have planets than their low activity counterparts.  This finding is especially evident in the unbiased sample of TESS TOIs which were chosen without prior knowledge of their activity levels. We also found an interesting bifurcation of the activity levels at late spectral types, with late-K and M stars showing separate populations of low and high activity stars. Finally, we present promising time series results that are consistent with previous work in G and K stars, and indicate potential cyclic behavior in new, lower mass targets.  All of the data obtained thus far will be made available on the OWLS website, and will be updated periodically as new data are obtained and analyzed.  

Our ultimate goal is to obtain activity cycle periods for a sizable sample of stars at each spectral type, and to produce a diagram such as Figure~1 of \citet{Bohm-Vitense2007} comparing these cycle periods to the rotation periods. Such data will provide valuable insight into the connection between the stellar rotation and the cyclic behavior of the magnetic dynamo, and how these properties change with the mass and internal structure of the star.  We also seek to further understand possible differences in the stellar magnetic activity in planet-host stars compared to those without planets.

\begin{acknowledgements}
    The MWO observations from \citet{HK_project} data derive from the Mount Wilson Observatory HK Project, which was supported by both public and private funds through the Carnegie Observatories, the Mount Wilson Institute, and the Harvard-Smithsonian Center for Astrophysics starting in 1966 and continuing for over 36 years. These data are the result of the dedicated work of O. Wilson, A. Vaughan, G. Preston, D. Duncan, S. Baliunas, and many others.  

    This research has made use of the NASA Exoplanet Archive, which is operated by the California Institute of Technology, under contract with the National Aeronautics and Space Administration under the Exoplanet Exploration Program

    We gratefully acknowledge the open source software which made this work possible: \texttt{astropy} \citep{Astropy2013, Astropy2018, AstropyCollaboration2022}, \texttt{astroquery} \citep{Ginsburg2019}, \texttt{specutils} \citep{specutils}, \texttt{ipython} \citep{ipython}, \texttt{numpy} \citep{numpy}, \texttt{scipy} \citep{scipy}, \texttt{matplotlib} \citep{matplotlib}, \texttt{astroplan} \citep{Morris2018astroplan}, \texttt{arcesetc} \citep{Morris2019c}, \texttt{aesop} \citep{Morris2018aesop}

\end{acknowledgements}

\facility{ARC}

\bibliographystyle{aasjournal} 
\bibliography{bibliography}

\begin{thebibliography}{}
\expandafter\ifx\csname natexlab\endcsname\relax\def\natexlab#1{#1}\fi
\providecommand{\url}[1]{\href{#1}{#1}}
\providecommand{\dodoi}[1]{doi:~\href{http://doi.org/#1}{\nolinkurl{#1}}}
\providecommand{\doeprint}[1]{\href{http://ascl.net/#1}{\nolinkurl{http://ascl.net/#1}}}
\providecommand{\doarXiv}[1]{\href{https://arxiv.org/abs/#1}{\nolinkurl{https://arxiv.org/abs/#1}}}

\bibitem[{{Airapetian} {et~al.}(2017){Airapetian}, {Glocer}, {Khazanov},
  {Loyd}, {France}, {Sojka}, {Danchi}, \& {Liemohn}}]{Airapetian2017}
{Airapetian}, V.~S., {Glocer}, A., {Khazanov}, G.~V., {et~al.} 2017, \apjl,
  836, L3, \dodoi{10.3847/2041-8213/836/1/L3}

\bibitem[{{Arney}(2019)}]{Arney2019}
{Arney}, G.~N. 2019, \apjl, 873, L7, \dodoi{10.3847/2041-8213/ab0651}

\bibitem[{{Arriagada}(2011)}]{Arriagada2011}
{Arriagada}, P. 2011, \apj, 734, 70, \dodoi{10.1088/0004-637X/734/1/70}

\bibitem[{{Astropy Collaboration} {et~al.}(2013){Astropy Collaboration},
  {Robitaille}, {Tollerud}, {Greenfield}, {Droettboom}, {Bray}, {Aldcroft},
  {Davis}, {Ginsburg}, {Price-Whelan}, {Kerzendorf}, {Conley}, {Crighton},
  {Barbary}, {Muna}, {Ferguson}, {Grollier}, {Parikh}, {Nair}, {Unther},
  {Deil}, {Woillez}, {Conseil}, {Kramer}, {Turner}, {Singer}, {Fox}, {Weaver},
  {Zabalza}, {Edwards}, {Azalee Bostroem}, {Burke}, {Casey}, {Crawford},
  {Dencheva}, {Ely}, {Jenness}, {Labrie}, {Lim}, {Pierfederici}, {Pontzen},
  {Ptak}, {Refsdal}, {Servillat}, \& {Streicher}}]{Astropy2013}
{Astropy Collaboration}, {Robitaille}, T.~P., {Tollerud}, E.~J., {et~al.} 2013,
  \aap, 558, A33, \dodoi{10.1051/0004-6361/201322068}

\bibitem[{{Astropy Collaboration} {et~al.}(2018){Astropy Collaboration},
  {Price-Whelan}, {Sip{\H{o}}cz}, {G{\"u}nther}, {Lim}, {Crawford}, {Conseil},
  {Shupe}, {Craig}, {Dencheva}, {Ginsburg}, {Vand erPlas}, {Bradley},
  {P{\'e}rez-Su{\'a}rez}, {de Val-Borro}, {Aldcroft}, {Cruz}, {Robitaille},
  {Tollerud}, {Ardelean}, {Babej}, {Bach}, {Bachetti}, {Bakanov}, {Bamford},
  {Barentsen}, {Barmby}, {Baumbach}, {Berry}, {Biscani}, {Boquien}, {Bostroem},
  {Bouma}, {Brammer}, {Bray}, {Breytenbach}, {Buddelmeijer}, {Burke},
  {Calderone}, {Cano Rodr{\'\i}guez}, {Cara}, {Cardoso}, {Cheedella}, {Copin},
  {Corrales}, {Crichton}, {D'Avella}, {Deil}, {Depagne}, {Dietrich}, {Donath},
  {Droettboom}, {Earl}, {Erben}, {Fabbro}, {Ferreira}, {Finethy}, {Fox},
  {Garrison}, {Gibbons}, {Goldstein}, {Gommers}, {Greco}, {Greenfield},
  {Groener}, {Grollier}, {Hagen}, {Hirst}, {Homeier}, {Horton}, {Hosseinzadeh},
  {Hu}, {Hunkeler}, {Ivezi{\'c}}, {Jain}, {Jenness}, {Kanarek}, {Kendrew},
  {Kern}, {Kerzendorf}, {Khvalko}, {King}, {Kirkby}, {Kulkarni}, {Kumar},
  {Lee}, {Lenz}, {Littlefair}, {Ma}, {Macleod}, {Mastropietro}, {McCully},
  {Montagnac}, {Morris}, {Mueller}, {Mumford}, {Muna}, {Murphy}, {Nelson},
  {Nguyen}, {Ninan}, {N{\"o}the}, {Ogaz}, {Oh}, {Parejko}, {Parley}, {Pascual},
  {Patil}, {Patil}, {Plunkett}, {Prochaska}, {Rastogi}, {Reddy Janga},
  {Sabater}, {Sakurikar}, {Seifert}, {Sherbert}, {Sherwood-Taylor}, {Shih},
  {Sick}, {Silbiger}, {Singanamalla}, {Singer}, {Sladen}, {Sooley},
  {Sornarajah}, {Streicher}, {Teuben}, {Thomas}, {Tremblay}, {Turner},
  {Terr{\'o}n}, {van Kerkwijk}, {de la Vega}, {Watkins}, {Weaver}, {Whitmore},
  {Woillez}, {Zabalza}, \& {Astropy Contributors}}]{Astropy2018}
{Astropy Collaboration}, {Price-Whelan}, A.~M., {Sip{\H{o}}cz}, B.~M., {et~al.}
  2018, \aj, 156, 123, \dodoi{10.3847/1538-3881/aabc4f}

\bibitem[{{Astropy Collaboration} {et~al.}(2022){Astropy Collaboration},
  {Price-Whelan}, {Lim}, {Earl}, {Starkman}, {Bradley}, {Shupe}, {Patil},
  {Corrales}, {Brasseur}, {N{\"o}the}, {Donath}, {Tollerud}, {Morris},
  {Ginsburg}, {Vaher}, {Weaver}, {Tocknell}, {Jamieson}, {van Kerkwijk},
  {Robitaille}, {Merry}, {Bachetti}, {G{\"u}nther}, {Aldcroft},
  {Alvarado-Montes}, {Archibald}, {B{\'o}di}, {Bapat}, {Barentsen},
  {Baz{\'a}n}, {Biswas}, {Boquien}, {Burke}, {Cara}, {Cara}, {Conroy},
  {Conseil}, {Craig}, {Cross}, {Cruz}, {D'Eugenio}, {Dencheva}, {Devillepoix},
  {Dietrich}, {Eigenbrot}, {Erben}, {Ferreira}, {Foreman-Mackey}, {Fox},
  {Freij}, {Garg}, {Geda}, {Glattly}, {Gondhalekar}, {Gordon}, {Grant},
  {Greenfield}, {Groener}, {Guest}, {Gurovich}, {Handberg}, {Hart},
  {Hatfield-Dodds}, {Homeier}, {Hosseinzadeh}, {Jenness}, {Jones}, {Joseph},
  {Kalmbach}, {Karamehmetoglu}, {Ka{\l}uszy{\'n}ski}, {Kelley}, {Kern},
  {Kerzendorf}, {Koch}, {Kulumani}, {Lee}, {Ly}, {Ma}, {MacBride}, {Maljaars},
  {Muna}, {Murphy}, {Norman}, {O'Steen}, {Oman}, {Pacifici}, {Pascual},
  {Pascual-Granado}, {Patil}, {Perren}, {Pickering}, {Rastogi}, {Roulston},
  {Ryan}, {Rykoff}, {Sabater}, {Sakurikar}, {Salgado}, {Sanghi}, {Saunders},
  {Savchenko}, {Schwardt}, {Seifert-Eckert}, {Shih}, {Jain}, {Shukla}, {Sick},
  {Simpson}, {Singanamalla}, {Singer}, {Singhal}, {Sinha}, {Sip{\H{o}}cz},
  {Spitler}, {Stansby}, {Streicher}, {{\v{S}}umak}, {Swinbank}, {Taranu},
  {Tewary}, {Tremblay}, {de Val-Borro}, {Van Kooten}, {Vasovi{\'c}}, {Verma},
  {de Miranda Cardoso}, {Williams}, {Wilson}, {Winkel}, {Wood-Vasey}, {Xue},
  {Yoachim}, {Zhang}, {Zonca}, \& {Astropy Project
  Contributors}}]{AstropyCollaboration2022}
{Astropy Collaboration}, {Price-Whelan}, A.~M., {Lim}, P.~L., {et~al.} 2022,
  The Astrophysical Journal, 935, 167, \dodoi{10.3847/1538-4357/ac7c74}

\bibitem[{{Astropy-Specutils Development Team}(2019)}]{specutils}
{Astropy-Specutils Development Team}. 2019, {Specutils: Spectroscopic analysis
  and reduction}.
\newblock \doeprint{1902.012}

\bibitem[{{Bakos} {et~al.}(2004){Bakos}, {Noyes}, {Kov{\'a}cs}, {Stanek},
  {Sasselov}, \& {Domsa}}]{bakos2004}
{Bakos}, G., {Noyes}, R.~W., {Kov{\'a}cs}, G., {et~al.} 2004, \pasp, 116, 266,
  \dodoi{10.1086/382735}

\bibitem[{{Bakos} {et~al.}(2010){Bakos}, {Torres}, {P{\'a}l}, {Hartman},
  {Kov{\'a}cs}, {Noyes}, {Latham}, {Sasselov}, {Sip{\H o}cz}, {Esquerdo},
  {Fischer}, {Johnson}, {Marcy}, {Butler}, {Isaacson}, {Howard}, {Vogt},
  {Kov{\'a}cs}, {Fernandez}, {Mo{\'o}r}, {Stefanik}, {L{\'a}z{\'a}r}, {Papp},
  \& {S{\'a}ri}}]{bakos2010}
{Bakos}, G.~{\'A}., {Torres}, G., {P{\'a}l}, A., {et~al.} 2010, \apj, 710,
  1724, \dodoi{10.1088/0004-637X/710/2/1724}

\bibitem[{{Baliunas} {et~al.}(1995){Baliunas}, {Donahue}, {Soon}, {Horne},
  {Frazer}, {Woodard-Eklund}, {Bradford}, {Rao}, {Wilson}, {Zhang}, {Bennett},
  {Briggs}, {Carroll}, {Duncan}, {Figueroa}, {Lanning}, {Misch}, {Mueller},
  {Noyes}, {Poppe}, {Porter}, {Robinson}, {Russell}, {Shelton}, {Soyumer},
  {Vaughan}, \& {Whitney}}]{Baliunas1995}
{Baliunas}, S.~L., {Donahue}, R.~A., {Soon}, W.~H., {et~al.} 1995, \apj, 438,
  269, \dodoi{10.1086/175072}

\bibitem[{{Barnes}(2007)}]{Barnes2007}
{Barnes}, S.~A. 2007, The Astrophysical Journal, 669, 1167,
  \dodoi{10.1086/519295}

\bibitem[{{B{\"o}hm-Vitense}(2007)}]{Bohm-Vitense2007}
{B{\"o}hm-Vitense}, E. 2007, The Astrophysical Journal, 657, 486,
  \dodoi{10.1086/510482}

\bibitem[{{Boro Saikia} {et~al.}(2018){Boro Saikia}, {Marvin}, {Jeffers},
  {Reiners}, {Cameron}, {Marsden}, {Petit}, {Warnecke}, \&
  {Yadav}}]{BoroSaikia2018}
{Boro Saikia}, S., {Marvin}, C.~J., {Jeffers}, S.~V., {et~al.} 2018, Astronomy
  and Astrophysics, 616, A108, \dodoi{10.1051/0004-6361/201629518}

\bibitem[{{Borucki} {et~al.}(2010){Borucki}, {Koch}, {Basri}, {Batalha},
  {Brown}, {Caldwell}, {Caldwell}, {Christensen-Dalsgaard}, {Cochran},
  {DeVore}, {Dunham}, {Dupree}, {Gautier}, {Geary}, {Gilliland}, {Gould},
  {Howell}, {Jenkins}, {Kondo}, {Latham}, {Marcy}, {Meibom}, {Kjeldsen},
  {Lissauer}, {Monet}, {Morrison}, {Sasselov}, {Tarter}, {Boss}, {Brownlee},
  {Owen}, {Buzasi}, {Charbonneau}, {Doyle}, {Fortney}, {Ford}, {Holman},
  {Seager}, {Steffen}, {Welsh}, {Rowe}, {Anderson}, {Buchhave}, {Ciardi},
  {Walkowicz}, {Sherry}, {Horch}, {Isaacson}, {Everett}, {Fischer}, {Torres},
  {Johnson}, {Endl}, {MacQueen}, {Bryson}, {Dotson}, {Haas}, {Kolodziejczak},
  {Van Cleve}, {Chandrasekaran}, {Twicken}, {Quintana}, {Clarke}, {Allen},
  {Li}, {Wu}, {Tenenbaum}, {Verner}, {Bruhweiler}, {Barnes}, \&
  {Prsa}}]{Kepler2010}
{Borucki}, W.~J., {Koch}, D., {Basri}, G., {et~al.} 2010, Science, 327, 977,
  \dodoi{10.1126/science.1185402}

\bibitem[{{Casagrande} \& {VandenBerg}(2018)}]{Casagrande2018}
{Casagrande}, L., \& {VandenBerg}, D.~A. 2018, \mnras, 479, L102,
  \dodoi{10.1093/mnrasl/sly104}

\bibitem[{{Cretignier} {et~al.}(2024){Cretignier}, {Pietrow}, \&
  {Aigrain}}]{Cretignier2024}
{Cretignier}, M., {Pietrow}, A.~G.~M., \& {Aigrain}, S. 2024, \mnras, 527,
  2940, \dodoi{10.1093/mnras/stad3292}

\bibitem[{{Cuntz} \& {Guinan}(2016)}]{Cuntz2016}
{Cuntz}, M., \& {Guinan}, E.~F. 2016, The Astrophysical Journal, 827, 79,
  \dodoi{10.3847/0004-637X/827/1/79}

\bibitem[{{Duncan} {et~al.}(1991){Duncan}, {Vaughan}, {Wilson}, {Preston},
  {Frazer}, {Lanning}, {Misch}, {Mueller}, {Soyumer}, {Woodard}, {Baliunas},
  {Noyes}, {Hartmann}, {Porter}, {Zwaan}, {Middelkoop}, {Rutten}, \&
  {Mihalas}}]{Duncan1991}
{Duncan}, D.~K., {Vaughan}, A.~H., {Wilson}, O.~C., {et~al.} 1991, \apjs, 76,
  383, \dodoi{10.1086/191572}

\bibitem[{{Egeland} {et~al.}(2017){Egeland}, {Soon}, {Baliunas}, {Hall},
  {Pevtsov}, \& {Bertello}}]{Egeland2017}
{Egeland}, R., {Soon}, W., {Baliunas}, S., {et~al.} 2017, \apj, 835, 25,
  \dodoi{10.3847/1538-4357/835/1/25}

\bibitem[{{Engle} \& {Guinan}(2023)}]{Engle2023}
{Engle}, S.~G., \& {Guinan}, E.~F. 2023, \apjl, 954, L50,
  \dodoi{10.3847/2041-8213/acf472}

\bibitem[{{Foreman-Mackey} {et~al.}(2013){Foreman-Mackey}, {Hogg}, {Lang}, \&
  {Goodman}}]{Foreman-Mackey2013}
{Foreman-Mackey}, D., {Hogg}, D.~W., {Lang}, D., \& {Goodman}, J. 2013,
  Publications of the Astronomical Society of the Pacific, 125, 306,
  \dodoi{10.1086/670067}

\bibitem[{{Fraunhofer}(1817)}]{Fraunhofer1817}
{Fraunhofer}, J. 1817, Annalen der Physik, 56, 264,
  \dodoi{10.1002/andp.18170560706}

\bibitem[{{Fuhrmeister} {et~al.}(2023){Fuhrmeister}, {Czesla}, {Perdelwitz},
  {Nagel}, {Schmitt}, {Jeffers}, {Caballero}, {Zechmeister}, {Montes},
  {Reiners}, {L{\'o}pez-Gallifa}, {Ribas}, {Quirrenbach}, {Amado},
  {Galad{\'\i}-Enr{\'\i}quez}, {B{\'e}jar}, {Danielski}, {Hatzes}, {Kaminski},
  {K{\"u}rster}, {Morales}, \& {Zapatero Osorio}}]{Fuhrmeister2023}
{Fuhrmeister}, B., {Czesla}, S., {Perdelwitz}, V., {et~al.} 2023, \aap, 670,
  A71, \dodoi{10.1051/0004-6361/202244829}

\bibitem[{{Gaia Collaboration} {et~al.}(2023){Gaia Collaboration}, {Vallenari},
  {Brown}, {Prusti}, {de Bruijne}, {Arenou}, {Babusiaux}, {Biermann},
  {Creevey}, {Ducourant}, {Evans}, {Eyer}, {Guerra}, {Hutton}, {Jordi},
  {Klioner}, {Lammers}, {Lindegren}, {Luri}, {Mignard}, {Panem}, {Pourbaix},
  {Randich}, {Sartoretti}, {Soubiran}, {Tanga}, {Walton}, {Bailer-Jones},
  {Bastian}, {Drimmel}, {Jansen}, {Katz}, {Lattanzi}, {van Leeuwen}, {Bakker},
  {Cacciari}, {Casta{\~n}eda}, {De Angeli}, {Fabricius}, {Fouesneau},
  {Fr{\'e}mat}, {Galluccio}, {Guerrier}, {Heiter}, {Masana}, {Messineo},
  {Mowlavi}, {Nicolas}, {Nienartowicz}, {Pailler}, {Panuzzo}, {Riclet}, {Roux},
  {Seabroke}, {Sordo}, {Th{\'e}venin}, {Gracia-Abril}, {Portell}, {Teyssier},
  {Altmann}, {Andrae}, {Audard}, {Bellas-Velidis}, {Benson}, {Berthier},
  {Blomme}, {Burgess}, {Busonero}, {Busso}, {C{\'a}novas}, {Carry}, {Cellino},
  {Cheek}, {Clementini}, {Damerdji}, {Davidson}, {de Teodoro}, {Nu{\~n}ez
  Campos}, {Delchambre}, {Dell'Oro}, {Esquej}, {Fern{\'a}ndez-Hern{\'a}ndez},
  {Fraile}, {Garabato}, {Garc{\'\i}a-Lario}, {Gosset}, {Haigron}, {Halbwachs},
  {Hambly}, {Harrison}, {Hern{\'a}ndez}, {Hestroffer}, {Hodgkin}, {Holl},
  {Jan{\ss}en}, {Jevardat de Fombelle}, {Jordan}, {Krone-Martins}, {Lanzafame},
  {L{\"o}ffler}, {Marchal}, {Marrese}, {Moitinho}, {Muinonen}, {Osborne},
  {Pancino}, {Pauwels}, {Recio-Blanco}, {Reyl{\'e}}, {Riello}, {Rimoldini},
  {Roegiers}, {Rybizki}, {Sarro}, {Siopis}, {Smith}, {Sozzetti}, {Utrilla},
  {van Leeuwen}, {Abbas}, {{\'A}brah{\'a}m}, {Abreu Aramburu}, {Aerts},
  {Aguado}, {Ajaj}, {Aldea-Montero}, {Altavilla}, {{\'A}lvarez}, {Alves},
  {Anders}, {Anderson}, {Anglada Varela}, {Antoja}, {Baines}, {Baker},
  {Balaguer-N{\'u}{\~n}ez}, {Balbinot}, {Balog}, {Barache}, {Barbato},
  {Barros}, {Barstow}, {Bartolom{\'e}}, {Bassilana}, {Bauchet}, {Becciani},
  {Bellazzini}, {Berihuete}, {Bernet}, {Bertone}, {Bianchi}, {Binnenfeld},
  {Blanco-Cuaresma}, {Blazere}, {Boch}, {Bombrun}, {Bossini}, {Bouquillon},
  {Bragaglia}, {Bramante}, {Breedt}, {Bressan}, {Brouillet}, {Brugaletta},
  {Bucciarelli}, {Burlacu}, {Butkevich}, {Buzzi}, {Caffau}, {Cancelliere},
  {Cantat-Gaudin}, {Carballo}, {Carlucci}, {Carnerero}, {Carrasco},
  {Casamiquela}, {Castellani}, {Castro-Ginard}, {Chaoul}, {Charlot}, {Chemin},
  {Chiaramida}, {Chiavassa}, {Chornay}, {Comoretto}, {Contursi}, {Cooper},
  {Cornez}, {Cowell}, {Crifo}, {Cropper}, {Crosta}, {Crowley}, {Dafonte},
  {Dapergolas}, {David}, {David}, {de Laverny}, {De Luise}, \& {De
  March}}]{gaiaDR3}
{Gaia Collaboration}, {Vallenari}, A., {Brown}, A.~G.~A., {et~al.} 2023,
  Astronomy and Astrophysics, 674, A1, \dodoi{10.1051/0004-6361/202243940}

\bibitem[{{Garcia-Sage} {et~al.}(2017){Garcia-Sage}, {Glocer}, {Drake},
  {Gronoff}, \& {Cohen}}]{GarciaSage2017}
{Garcia-Sage}, K., {Glocer}, A., {Drake}, J.~J., {Gronoff}, G., \& {Cohen}, O.
  2017, \apjl, 844, L13, \dodoi{10.3847/2041-8213/aa7eca}

\bibitem[{{Ginsburg} {et~al.}(2019){Ginsburg}, {Sip{\H{o}}cz}, {Brasseur},
  {Cowperthwaite}, {Craig}, {Deil}, {Guillochon}, {Guzman}, {Liedtke}, {Lian
  Lim}, {Lockhart}, {Mommert}, {Morris}, {Norman}, {Parikh}, {Persson},
  {Robitaille}, {Segovia}, {Singer}, {Tollerud}, {de Val-Borro}, {Valtchanov},
  {Woillez}, {Astroquery Collaboration}, \& {a subset of astropy
  Collaboration}}]{Ginsburg2019}
{Ginsburg}, A., {Sip{\H{o}}cz}, B.~M., {Brasseur}, C.~E., {et~al.} 2019, The
  Astronomical Journal, 157, 98, \dodoi{10.3847/1538-3881/aafc33}

\bibitem[{{Gomes da Silva} {et~al.}(2021){Gomes da Silva}, {Santos},
  {Adibekyan}, {Sousa}, {Campante}, {Figueira}, {Bossini}, {Delgado-Mena},
  {Monteiro}, {de Laverny}, {Recio-Blanco}, \& {Lovis}}]{GomesdaSilva2021}
{Gomes da Silva}, J., {Santos}, N.~C., {Adibekyan}, V., {et~al.} 2021, \aap,
  646, A77, \dodoi{10.1051/0004-6361/202039765}

\bibitem[{{Goodman} \& {Weare}(2010)}]{Goodman2010}
{Goodman}, J., \& {Weare}, J. 2010, Communications in Applied Mathematics and
  Computational Science, 5, 65, \dodoi{10.2140/camcos.2010.5.65}

\bibitem[{{Guerrero} {et~al.}(2021){Guerrero}, {Seager}, {Huang}, {Vanderburg},
  {Garcia Soto}, {Mireles}, {Hesse}, {Fong}, {Glidden}, {Shporer}, {Latham},
  {Collins}, {Quinn}, {Burt}, {Dragomir}, {Crossfield}, {Vanderspek},
  {Fausnaugh}, {Burke}, {Ricker}, {Daylan}, {Essack}, {G{\"u}nther}, {Osborn},
  {Pepper}, {Rowden}, {Sha}, {Villanueva}, {Yahalomi}, {Yu}, {Ballard},
  {Batalha}, {Berardo}, {Chontos}, {Dittmann}, {Esquerdo}, {Mikal-Evans},
  {Jayaraman}, {Krishnamurthy}, {Louie}, {Mehrle}, {Niraula}, {Rackham},
  {Rodriguez}, {Rowden}, {Sousa-Silva}, {Watanabe}, {Wong}, {Zhan},
  {Zivanovic}, {Christiansen}, {Ciardi}, {Swain}, {Lund}, {Mullally},
  {Fleming}, {Rodriguez}, {Boyd}, {Quintana}, {Barclay}, {Col{\'o}n},
  {Rinehart}, {Schlieder}, {Clampin}, {Jenkins}, {Twicken}, {Caldwell},
  {Coughlin}, {Henze}, {Lissauer}, {Morris}, {Rose}, {Smith}, {Tenenbaum},
  {Ting}, {Wohler}, {Bakos}, {Bean}, {Berta-Thompson}, {Bieryla}, {Bouma},
  {Buchhave}, {Butler}, {Charbonneau}, {Doty}, {Ge}, {Holman}, {Howard},
  {Kaltenegger}, {Kane}, {Kjeldsen}, {Kreidberg}, {Lin}, {Minsky}, {Narita},
  {Paegert}, {P{\'a}l}, {Palle}, {Sasselov}, {Spencer}, {Sozzetti}, {Stassun},
  {Torres}, {Udry}, \& {Winn}}]{TOIs2021}
{Guerrero}, N.~M., {Seager}, S., {Huang}, C.~X., {et~al.} 2021, \apjs, 254, 39,
  \dodoi{10.3847/1538-4365/abefe1}

\bibitem[{{Hall} {et~al.}(2007){Hall}, {Lockwood}, \& {Skiff}}]{Hall2007}
{Hall}, J.~C., {Lockwood}, G.~W., \& {Skiff}, B.~A. 2007, \aj, 133, 862,
  \dodoi{10.1086/510356}

\bibitem[{Harris {et~al.}(2020)Harris, Millman, van~der Walt, Gommers,
  Virtanen, Cournapeau, Wieser, Taylor, Berg, Smith, Kern, Picus, Hoyer, van
  Kerkwijk, Brett, Haldane, del R{\'{i}}o, Wiebe, Peterson,
  G{\'{e}}rard-Marchant, Sheppard, Reddy, Weckesser, Abbasi, Gohlke, \&
  Oliphant}]{numpy}
Harris, C.~R., Millman, K.~J., van~der Walt, S.~J., {et~al.} 2020, Nature, 585,
  357, \dodoi{10.1038/s41586-020-2649-2}

\bibitem[{{Hatalova} {et~al.}(2023){Hatalova}, {Brasser}, {Mamonova}, \&
  {Werner}}]{Hatalova2023}
{Hatalova}, P., {Brasser}, R., {Mamonova}, E., \& {Werner}, S.~C. 2023, \aap,
  676, A131, \dodoi{10.1051/0004-6361/202346332}

\bibitem[{{Hawley} {et~al.}(1996){Hawley}, {Gizis}, \& {Reid}}]{PMSU2}
{Hawley}, S.~L., {Gizis}, J.~E., \& {Reid}, I.~N. 1996, \aj, 112, 2799,
  \dodoi{10.1086/118222}

\bibitem[{{Hunter}(2007)}]{matplotlib}
{Hunter}, J.~D. 2007, Computing in Science and Engineering, 9, 90,
  \dodoi{10.1109/MCSE.2007.55}

\bibitem[{{Husser} {et~al.}(2013){Husser}, {Wende-von Berg}, {Dreizler},
  {Homeier}, {Reiners}, {Barman}, \& {Hauschildt}}]{Husser2013}
{Husser}, T.-O., {Wende-von Berg}, S., {Dreizler}, S., {et~al.} 2013, \aap,
  553, A6, \dodoi{10.1051/0004-6361/201219058}

\bibitem[{{Isaacson} \& {Fischer}(2010)}]{Isaacson2010}
{Isaacson}, H., \& {Fischer}, D. 2010, \apj, 725, 875,
  \dodoi{10.1088/0004-637X/725/1/875}

\bibitem[{{Isaacson} {et~al.}(2024){Isaacson}, {Howard}, {Fulton}, {Petigura},
  {Weiss}, {Kane}, {Carter}, {Beard}, {Giacalone}, {Van Zandt}, {Murphy},
  {Dai}, {Chontos}, {Polanski}, {Rice}, {Lubin}, {Brinkman}, {Rubenzahl},
  {Blunt}, {Yee}, {MacDougall}, {Dalba}, {Tyler}, {Behmard}, {Angelo},
  {Pidhorodetska}, {Mayo}, {Holcomb}, {Turtelboom}, {Hill}, {Bouma}, {Zhang},
  {Crossfield}, \& {Saunders}}]{Isaacson2024}
{Isaacson}, H., {Howard}, A.~W., {Fulton}, B., {et~al.} 2024, \apjs, 274, 35,
  \dodoi{10.3847/1538-4365/ad676c}

\bibitem[{{Kruse} {et~al.}(2019){Kruse}, {Agol}, {Luger}, \&
  {Foreman-Mackey}}]{Kruse2019}
{Kruse}, E., {Agol}, E., {Luger}, R., \& {Foreman-Mackey}, D. 2019, \apjs, 244,
  11, \dodoi{10.3847/1538-4365/ab346b}

\bibitem[{{Landolt} \& {Uomoto}(2007)}]{Landolt2007}
{Landolt}, A.~U., \& {Uomoto}, A.~K. 2007, \aj, 133, 768,
  \dodoi{10.1086/510485}

\bibitem[{{Leighton}(1959)}]{Leighton1959}
{Leighton}, R.~B. 1959, \apj, 130, 366, \dodoi{10.1086/146727}

\bibitem[{{Lillo-Box} {et~al.}(2022){Lillo-Box}, {Santos}, {Santerne}, {Silva},
  {Barrado}, {Faria}, {Castro-Gonz{\'a}lez}, {Balsalobre-Ruza},
  {Morales-Calder{\'o}n}, {Saavedra}, {Marfil}, {Sousa}, {Adibekyan},
  {Berihuete}, {Barros}, {Delgado-Mena}, {Hu{\'e}lamo}, {Deleuil}, {Demangeon},
  {Figueira}, {Grouffal}, {Aceituno}, {Azzaro}, {Bergond},
  {Fern{\'a}ndez-Mart{\'\i}n}, {Galad{\'\i}}, {Gallego}, {Gardini},
  {G{\'o}ngora}, {Guijarro}, {Hermelo}, {Mart{\'\i}n}, {M{\'\i}nguez},
  {Montoya}, {Pedraz}, \& {Vico Linares}}]{LilloBox2022}
{Lillo-Box}, J., {Santos}, N.~C., {Santerne}, A., {et~al.} 2022, \aap, 667,
  A102, \dodoi{10.1051/0004-6361/202243898}

\bibitem[{{Linsky} {et~al.}(1979){Linsky}, {Worden}, {McClintock}, \&
  {Robertson}}]{Linsky1979}
{Linsky}, J.~L., {Worden}, S.~P., {McClintock}, W., \& {Robertson}, R.~M. 1979,
  \apjs, 41, 47, \dodoi{10.1086/190607}

\bibitem[{{Lorenzo-Oliveira} {et~al.}(2018){Lorenzo-Oliveira}, {Freitas},
  {Mel{\'e}ndez}, {Bedell}, {Ram{\'\i}rez}, {Bean}, {Asplund}, {Spina},
  {Dreizler}, {Alves-Brito}, \& {Casagrande}}]{Lorenzo-Oliveira2018}
{Lorenzo-Oliveira}, D., {Freitas}, F.~C., {Mel{\'e}ndez}, J., {et~al.} 2018,
  Astronomy and Astrophysics, 619, A73, \dodoi{10.1051/0004-6361/201629294}

\bibitem[{{Lovis} {et~al.}(2011){Lovis}, {Dumusque}, {Santos}, {Bouchy},
  {Mayor}, {Pepe}, {Queloz}, {S{\'e}gransan}, \& {Udry}}]{Lovis2011}
{Lovis}, C., {Dumusque}, X., {Santos}, N.~C., {et~al.} 2011, ArXiv e-prints.
\newblock \doarXiv{1107.5325}

\bibitem[{{Luhn} {et~al.}(2020){Luhn}, {Wright}, {Howard}, \&
  {Isaacson}}]{Luhn2020}
{Luhn}, J.~K., {Wright}, J.~T., {Howard}, A.~W., \& {Isaacson}, H. 2020, \aj,
  159, 235, \dodoi{10.3847/1538-3881/ab855a}

\bibitem[{{Mamajek} \& {Hillenbrand}(2008)}]{Mamajek2008}
{Mamajek}, E.~E., \& {Hillenbrand}, L.~A. 2008, The Astrophysical Journal, 687,
  1264, \dodoi{10.1086/591785}

\bibitem[{{Marvin} {et~al.}(2023){Marvin}, {Reiners}, {Anglada-Escud{\'e}},
  {Jeffers}, \& {Boro Saikia}}]{Marvin2023}
{Marvin}, C.~J., {Reiners}, A., {Anglada-Escud{\'e}}, G., {Jeffers}, S.~V., \&
  {Boro Saikia}, S. 2023, \aap, 671, A162, \dodoi{10.1051/0004-6361/201937306}

\bibitem[{{Massey} {et~al.}(1988){Massey}, {Strobel}, {Barnes}, \&
  {Anderson}}]{Massey1988}
{Massey}, P., {Strobel}, K., {Barnes}, J.~V., \& {Anderson}, E. 1988, \apj,
  328, 315, \dodoi{10.1086/166294}

\bibitem[{{McQuillan} {et~al.}(2013){McQuillan}, {Mazeh}, \&
  {Aigrain}}]{McQuillan2013}
{McQuillan}, A., {Mazeh}, T., \& {Aigrain}, S. 2013, \apj, 775, L11,
  \dodoi{10.1088/2041-8205/775/1/L11}

\bibitem[{{Mignon} {et~al.}(2023){Mignon}, {Meunier}, {Delfosse}, {Bonfils},
  {Santos}, {Forveille}, {Gaisn{\'e}}, {Astudillo-Defru}, {Lovis}, \&
  {Udry}}]{Mignon2023}
{Mignon}, L., {Meunier}, N., {Delfosse}, X., {et~al.} 2023, \aap, 675, A168,
  \dodoi{10.1051/0004-6361/202244249}

\bibitem[{{Mittag} {et~al.}(2013){Mittag}, {Schmitt}, \&
  {Schr{\"o}der}}]{Mittag2013}
{Mittag}, M., {Schmitt}, J.~H.~M.~M., \& {Schr{\"o}der}, K.-P. 2013, \aap, 549,
  A117, \dodoi{10.1051/0004-6361/201219868}

\bibitem[{{Morris} {et~al.}(2019{\natexlab{a}}){Morris}, {Curtis}, {Sakari},
  {Hawley}, \& {Agol}}]{Morris2019b}
{Morris}, B.~M., {Curtis}, J.~L., {Sakari}, C., {Hawley}, S.~L., \& {Agol}, E.
  2019{\natexlab{a}}, arXiv e-prints, arXiv:1907.00423.
\newblock \doarXiv{1907.00423}

\bibitem[{{Morris} {et~al.}(2019{\natexlab{b}}){Morris}, {Davenport}, {Giles},
  {Hebb}, {Hawley}, {Angus}, {Gilman}, \& {Agol}}]{Morris2019c}
{Morris}, B.~M., {Davenport}, J. R.~A., {Giles}, H. A.~C., {et~al.}
  2019{\natexlab{b}}, Monthly Notices of the Royal Astronomical Society, 484,
  3244, \dodoi{10.1093/mnras/stz199}

\bibitem[{{Morris} \& {Dorn-Wallenstein}(2018)}]{Morris2018aesop}
{Morris}, B.~M., \& {Dorn-Wallenstein}, T. 2018, The Journal of Open Source
  Software, 3, 854, \dodoi{10.21105/joss.00854}

\bibitem[{{Morris} {et~al.}(2017){Morris}, {Hawley}, {Hebb}, {Sakari},
  {Davenport}, {Isaacson}, {Howard}, {Montet}, \& {Agol}}]{Morris2017b}
{Morris}, B.~M., {Hawley}, S.~L., {Hebb}, L., {et~al.} 2017, \apj, 848, 58,
  \dodoi{10.3847/1538-4357/aa8cca}

\bibitem[{{Morris} {et~al.}(2018){Morris}, {Tollerud}, {Sip{\H{o}}cz}, {Deil},
  {Douglas}, {Berlanga Medina}, {Vyhmeister}, {Smith}, {Littlefair},
  {Price-Whelan}, {Gee}, \& {Jeschke}}]{Morris2018astroplan}
{Morris}, B.~M., {Tollerud}, E., {Sip{\H{o}}cz}, B., {et~al.} 2018, \aj, 155,
  128, \dodoi{10.3847/1538-3881/aaa47e}

\bibitem[{{Noyes} {et~al.}(1984){Noyes}, {Hartmann}, {Baliunas}, {Duncan}, \&
  {Vaughan}}]{Noyes1984}
{Noyes}, R.~W., {Hartmann}, L.~W., {Baliunas}, S.~L., {Duncan}, D.~K., \&
  {Vaughan}, A.~H. 1984, \apj, 279, 763, \dodoi{10.1086/161945}

\bibitem[{{Oke}(1990)}]{Oke1990}
{Oke}, J.~B. 1990, \aj, 99, 1621, \dodoi{10.1086/115444}

\bibitem[{{Owen} \& {Mohanty}(2016)}]{Owen2016}
{Owen}, J.~E., \& {Mohanty}, S. 2016, \mnras, 459, 4088,
  \dodoi{10.1093/mnras/stw959}

\bibitem[{{Pecaut} \& {Mamajek}(2013)}]{Pecaut2013}
{Pecaut}, M.~J., \& {Mamajek}, E.~E. 2013, \apjs, 208, 9,
  \dodoi{10.1088/0067-0049/208/1/9}

\bibitem[{P\'erez \& Granger(2007)}]{ipython}
P\'erez, F., \& Granger, B.~E. 2007, Computing in Science and Engineering, 9,
  21, \dodoi{10.1109/MCSE.2007.53}

\bibitem[{{Pollacco} {et~al.}(2006){Pollacco}, {Skillen}, {Collier Cameron},
  {Christian}, {Hellier}, {Irwin}, {Lister}, {Street}, {West}, {Anderson},
  {Clarkson}, {Deeg}, {Enoch}, {Evans}, {Fitzsimmons}, {Haswell}, {Hodgkin},
  {Horne}, {Kane}, {Keenan}, {Maxted}, {Norton}, {Osborne}, {Parley}, {Ryans},
  {Smalley}, {Wheatley}, \& {Wilson}}]{Pollacco2006}
{Pollacco}, D.~L., {Skillen}, I., {Collier Cameron}, A., {et~al.} 2006, \pasp,
  118, 1407, \dodoi{10.1086/508556}

\bibitem[{Radick \& Pevtsov(2018)}]{HK_project}
Radick, R., \& Pevtsov, A. 2018, {HK\_Project\_v1995\_NSO}, V1,  Harvard
  Dataverse, \dodoi{10.7910/DVN/ZRJ6NT}

\bibitem[{{Radick} {et~al.}(1998){Radick}, {Lockwood}, {Skiff}, \&
  {Baliunas}}]{Radick1998}
{Radick}, R.~R., {Lockwood}, G.~W., {Skiff}, B.~A., \& {Baliunas}, S.~L. 1998,
  The Astrophysical Journal Supplement Series, 118, 239, \dodoi{10.1086/313135}

\bibitem[{{Reid} {et~al.}(1995){Reid}, {Hawley}, \& {Gizis}}]{PMSU1}
{Reid}, I.~N., {Hawley}, S.~L., \& {Gizis}, J.~E. 1995, \aj, 110, 1838,
  \dodoi{10.1086/117655}

\bibitem[{{Richey-Yowell} {et~al.}(2019){Richey-Yowell}, {Shkolnik},
  {Schneider}, {Osby}, {Barman}, \& {Meadows}}]{Richey-Yowell2019}
{Richey-Yowell}, T., {Shkolnik}, E.~L., {Schneider}, A.~C., {et~al.} 2019, The
  Astrophysical Journal, 872, 17, \dodoi{10.3847/1538-4357/aafa74}

\bibitem[{{Richey-Yowell} {et~al.}(2023){Richey-Yowell}, {Shkolnik},
  {Schneider}, {Peacock}, {Huseby}, {Jackman}, {Barman}, {Osby}, \&
  {Meadows}}]{Richey-Yowell2023}
---. 2023, The Astrophysical Journal, 951, 44, \dodoi{10.3847/1538-4357/acd2dc}

\bibitem[{{Ricker} {et~al.}(2015){Ricker}, {Winn}, {Vanderspek}, {Latham},
  {Bakos}, {Bean}, {Berta-Thompson}, {Brown}, {Buchhave}, {Butler}, {Butler},
  {Chaplin}, {Charbonneau}, {Christensen-Dalsgaard}, {Clampin}, {Deming},
  {Doty}, {De Lee}, {Dressing}, {Dunham}, {Endl}, {Fressin}, {Ge}, {Henning},
  {Holman}, {Howard}, {Ida}, {Jenkins}, {Jernigan}, {Johnson}, {Kaltenegger},
  {Kawai}, {Kjeldsen}, {Laughlin}, {Levine}, {Lin}, {Lissauer}, {MacQueen},
  {Marcy}, {McCullough}, {Morton}, {Narita}, {Paegert}, {Palle}, {Pepe},
  {Pepper}, {Quirrenbach}, {Rinehart}, {Sasselov}, {Sato}, {Seager},
  {Sozzetti}, {Stassun}, {Sullivan}, {Szentgyorgyi}, {Torres}, {Udry}, \&
  {Villasenor}}]{TESS2015}
{Ricker}, G.~R., {Winn}, J.~N., {Vanderspek}, R., {et~al.} 2015, Journal of
  Astronomical Telescopes, Instruments, and Systems, 1, 014003,
  \dodoi{10.1117/1.JATIS.1.1.014003}

\bibitem[{{Saar} \& {Osten}(1997)}]{Saar1997}
{Saar}, S.~H., \& {Osten}, R.~A. 1997, Monthly Notices of the Royal
  Astronomical Society, 284, 803, \dodoi{10.1093/mnras/284.4.803}

\bibitem[{{Schrijver}(1987)}]{Schrijver1987}
{Schrijver}, C.~J. 1987, \aap, 172, 111

\bibitem[{{Shkolnik} \& {Barman}(2014)}]{Shkolnik2014}
{Shkolnik}, E.~L., \& {Barman}, T.~S. 2014, \aj, 148, 64,
  \dodoi{10.1088/0004-6256/148/4/64}

\bibitem[{{Skumanich}(1972)}]{Skumanich1972}
{Skumanich}, A. 1972, \apj, 171, 565, \dodoi{10.1086/151310}

\bibitem[{{Solanki}(2003)}]{Solanki2003}
{Solanki}, S.~K. 2003, \aapr, 11, 153, \dodoi{10.1007/s00159-003-0018-4}

\bibitem[{{Thompson} {et~al.}(2018){Thompson}, {Coughlin}, {Hoffman},
  {Mullally}, {Christiansen}, {Burke}, {Bryson}, {Batalha}, {Haas},
  {Catanzarite}, {Rowe}, {Barentsen}, {Caldwell}, {Clarke}, {Jenkins}, {Li},
  {Latham}, {Lissauer}, {Mathur}, {Morris}, {Seader}, {Smith}, {Klaus},
  {Twicken}, {Van Cleve}, {Wohler}, {Akeson}, {Ciardi}, {Cochran}, {Henze},
  {Howell}, {Huber}, {Pr{\v{s}}a}, {Ram{\'\i}rez}, {Morton}, {Barclay},
  {Campbell}, {Chaplin}, {Charbonneau}, {Christensen-Dalsgaard}, {Dotson},
  {Doyle}, {Dunham}, {Dupree}, {Ford}, {Geary}, {Girouard}, {Isaacson},
  {Kjeldsen}, {Quintana}, {Ragozzine}, {Shabram}, {Shporer}, {Silva Aguirre},
  {Steffen}, {Still}, {Tenenbaum}, {Welsh}, {Wolfgang}, {Zamudio}, {Koch}, \&
  {Borucki}}]{KOIs2018}
{Thompson}, S.~E., {Coughlin}, J.~L., {Hoffman}, K., {et~al.} 2018, \apjs, 235,
  38, \dodoi{10.3847/1538-4365/aab4f9}

\bibitem[{{Vaughan} {et~al.}(1978){Vaughan}, {Preston}, \&
  {Wilson}}]{Vaughan1978}
{Vaughan}, A.~H., {Preston}, G.~W., \& {Wilson}, O.~C. 1978, \pasp, 90, 267,
  \dodoi{10.1086/130324}

\bibitem[{{Vilovi{\'c}} {et~al.}(2024){Vilovi{\'c}}, {Schulze-Makuch}, \&
  {Heller}}]{Vilovic2024}
{Vilovi{\'c}}, I., {Schulze-Makuch}, D., \& {Heller}, R. 2024, International
  Journal of Astrobiology, 23, e18, \dodoi{10.1017/S1473550424000132}

\bibitem[{Virtanen {et~al.}(2020)Virtanen, Gommers, Oliphant, Haberland, Reddy,
  Cournapeau, Burovski, Peterson, Weckesser, Bright, {van der Walt}, Brett,
  Wilson, Millman, Mayorov, Nelson, Jones, Kern, Larson, Carey, Polat, Feng,
  Moore, {VanderPlas}, Laxalde, Perktold, Cimrman, Henriksen, Quintero, Harris,
  Archibald, Ribeiro, Pedregosa, {van Mulbregt}, \& {SciPy 1.0
  Contributors}}]{scipy}
Virtanen, P., Gommers, R., Oliphant, T.~E., {et~al.} 2020, Nature Methods, 17,
  261, \dodoi{10.1038/s41592-019-0686-2}

\bibitem[{{Wilson} \& {Woolley}(1970)}]{Wilson1970}
{Wilson}, O., \& {Woolley}, R. 1970, \mnras, 148, 463,
  \dodoi{10.1093/mnras/148.4.463}

\bibitem[{{Wilson}(1963)}]{Wilson1963}
{Wilson}, O.~C. 1963, \apj, 138, 832, \dodoi{10.1086/147689}

\bibitem[{{Wilson}(1968)}]{Wilson1968}
---. 1968, \apj, 153, 221, \dodoi{10.1086/149652}

\bibitem[{{Wilson}(1976)}]{Wilson1976}
---. 1976, \apj, 205, 823, \dodoi{10.1086/154338}

\bibitem[{{Wilson}(1978)}]{Wilson1978}
---. 1978, \apj, 226, 379, \dodoi{10.1086/156618}

\bibitem[{{Wilson} \& {Skumanich}(1964)}]{Wilson1964}
{Wilson}, O.~C., \& {Skumanich}, A. 1964, \apj, 140, 1401,
  \dodoi{10.1086/148046}

\bibitem[{{Wilson} \& {Vainu Bappu}(1957)}]{Wilson1957}
{Wilson}, O.~C., \& {Vainu Bappu}, M.~K. 1957, \apj, 125, 661,
  \dodoi{10.1086/146339}

\bibitem[{{Wright} {et~al.}(2004){Wright}, {Marcy}, {Butler}, \&
  {Vogt}}]{Wright2004}
{Wright}, J.~T., {Marcy}, G.~W., {Butler}, R.~P., \& {Vogt}, S.~S. 2004, \apjs,
  152, 261, \dodoi{10.1086/386283}

\end{thebibliography}

\end{document}